\documentclass[pdftex,twocolumn,epjc3]{svjour3}          

\RequirePackage[T1]{fontenc}

\smartqed  

\RequirePackage{graphicx}
\RequirePackage{mathptmx}      
\RequirePackage{flushend}
\RequirePackage[numbers,sort&compress]{natbib}
\RequirePackage[colorlinks,citecolor=blue,urlcolor=blue,linkcolor=blue]{hyperref}

\journalname{Eur. Phys. J. C}

\usepackage{graphicx}
\usepackage{dcolumn}
\usepackage{bm}
\usepackage{units}
\usepackage{url}
\usepackage[mathlines]{lineno}
\usepackage{natbib}
\usepackage{varwidth}
\usepackage{color}
\usepackage{float}
\usepackage{comment}
\usepackage{lipsum}
\usepackage{graphicx}
\usepackage{lineno}
\usepackage{gensymb}
\usepackage[italic]{hepnames}
\usepackage{siunitx}
\usepackage{placeins}
\usepackage{footnote}
\makesavenoteenv{tabular}
\makesavenoteenv{table}

\usepackage{amsmath}
\usepackage{mathrsfs}

\usepackage{array, makecell} %

\DeclareFontFamily{U}{calligra}{}
\DeclareFontShape{U}{calligra}{m}{n}{<->callig15}{}

\begin{document}


\title{Predicting Transport Effects of Scintillation Light
  Signals in Large-Scale Liquid Argon Detectors}

  \author{Diego Garcia-Gamez\thanksref{e1, addr1} 
  \and Patrick Green\thanksref{addr2} 
  \and Andrzej M. Szelc\thanksref{e3, addr2}
  }
\thankstext{e1}{e-mail: dgarciag@ugr.es}
\thankstext{e3}{Now at University of Edinburgh}
\institute{University of Granada \& CAFPE, Campus Fuentenueva, 18002 Granada, Spain \label{addr1}
\and
Department of Physics and Astronomy, University of Manchester, Oxford Road, Manchester, M13 9PL, UK \label{addr2} 
}%


\maketitle


\begin{abstract}
Liquid argon is being employed as a detector medium in neutrino physics and Dark Matter searches. A recent push to expand the applications of scintillation light in Liquid Argon Time Projection Chamber neutrino detectors has necessitated the development of advanced methods of simulating this light. The presently available methods tend to be prohibitively slow or imprecise due to the combination of detector size and the amount of energy deposited by neutrino beam interactions. In this work we present a semi-analytical model to predict the quantity of argon scintillation light observed by a light detector with a precision better than $10\%$, based only on the relative positions between the scintillation and light detector. We also provide a method to predict the distribution of arrival times of these photons accounting for propagation effects. Additionally, we present an equivalent model to predict the number of photons and their arrival times in the case of a wavelength-shifting, highly-reflective layer being present on the detector cathode. Our proposed method can be used to simulate light propagation in large-scale liquid argon detectors such as DUNE or SBND, and could also be applied to other detector mediums such as liquid xenon or xenon-doped liquid argon.

\end{abstract}

\section{Introduction} \label{sec:introduction}

Several experiments setting out to search for the elusive dark matter particles~\cite{Benetti:2007cd, Aalseth:2017fik, AMAUDRUZ20191} or perform precision measurements of neutrino parameters~\cite{Acciarri:2016smi, Machado:2019oxb,Abi:2020evt} have chosen liquid argon (LAr) as their detector medium. Liquid argon is relatively dense ($1.41$\,g/cm$^3$) and is chemically inert allowing ionisation charge to be drifted for distances of several metres. These properties combined with its relatively low price make liquid argon an excellent choice for large scale detectors~\cite{Rubbia:1977zz}, enabling the construction of modules as large as several kTons~\cite{Abi:2020evt}. In addition to the ionisation charge used by Liquid Argon Time Projection Chamber (LArTPC) neutrino detectors, liquid argon is an excellent scintillator emitting on the order of 40000 photons per MeV of deposited energy. This scintillation light has been employed by experiments searching for dark matter for energy reconstruction and background rejection~\cite{Benetti:2007cd, Agnes:2014bvk, Ajaj:2019}. In neutrino detectors however, liquid argon scintillation light has not been as thoroughly exploited to date. The MicroBooNE~\cite{Acciarri:2016smi} and ICARUS~\cite{Amerio:2004ze} experiments have mostly used scintillation light as a means of triggering and rejecting cosmic events. It has been recently proposed~\cite{Sorel:2014rka,Acciarri:2019wgd,Szelc:2016rjm} that LArTPC neutrino detectors with enhanced light collection capabilities could employ scintillation light for improved time, calorimetry and position resolution. 

More sophisticated applications of scintillation light in LArTPC neutrino detectors will require precise simulation of the light to develop new algorithms and validate their performance. Such simulation quickly becomes computationally challenging when the detector size reaches tens of tons (or even kTons as in the case of DUNE) combined with the events of interest depositing hundreds of MeV of energy, as expected for accelerator neutrinos. Simulating each photon emitted by the interacting events individually becomes prohibitively slow. A solution applied in the LArSoft software package~\cite{Snider:2017wjd} used by most LArTPC detectors is to implement a lookup library method~\cite{Jones:2015bya}. In this method the computationally challenging work is performed once, by running a large stand-alone job that generates a vast number of isotropic photons from pre-defined cubes (voxels) inside of the detector active volume. For all voxel-photon-detector pairs the probability of light being detected is saved in a dedicated file. This file is then accessed by standard simulation jobs that use it to estimate the number of detected photons for a given energy deposition without having to simulate each photon separately. This approach works reasonably well in estimating the amount of light for detectors of tens of tons such as MicroBooNE or SBND, albeit introducing a certain granularity into the simulation driven by the size of the voxels used. However, for extremely large detectors this approach results in very large lookup files of several GB even when considering relatively large voxel sizes. Additionally, to develop applications of scintillation light involving time, a good understanding of the effects of light propagation in the medium is needed in addition to the understanding of the intrinsic scintillation light components~\cite{Doke:1988rq} and the effects of wavelength shifting, e.g. by tetraphenyl-butadiene (TPB)~\cite{Segreto:2014aia} or the recently proposed polyethylene naphthalate (PEN)~\cite{Kuzniak:2018dcf}. At the large distances present in currently built and proposed LArTPCs, Rayleigh scattering becomes an important factor. This scattering can result in a smearing of the photon arrival times leading to non-trivial effects in their arrival time distributions. The lookup-library method is not designed to predict these effects and, given the non-trivial distributions, a brute-force approach could necessitate storing a function definition (or a histogram) for each voxel-photon-detector pair. Incorporating this could greatly increase the size of the required lookup-library.  

We propose a method to numerically predict the number of photons observed in a particular photon-detector based only on the size and position of the energy depositions in the liquid argon volume. This approach allows fast simulation of scintillation light in large scale liquid argon detectors with a precision better than 10\%
, even for cases of large numbers of photons originating from high energy particle interactions. In addition, we have developed an analogous method to predict the number of detected photons arriving from a wavelength-shifter coated, highly-reflective, detector cathode. Passive light collection elements of this type, in the form of reflective TPB-coated foils, are planned to be installed in the SBND experiment~\cite{Basque2019_foils} and have been proposed as an option for the DUNE detectors~\cite{Abi2020_tdr_iv}. Finally, we also present a method to generate the distribution of photon arrival times that accounts for effects of Rayleigh scattering in liquid argon, and reflections and absorption on the detector walls, using only the relative positions of the energy deposition and the photon detectors. Though developed for the liquid argon case, these methods are in principle applicable to any medium with a refractive index such that the Rayleigh scattering length is comparable to the size of volume in which the light propagates. They could therefore be applied to liquid xenon or xenon-doped argon detectors. 

This article is structured as follows: we first present the basics of liquid argon scintillation light emission and detection, that are relevant to the simulation method we propose. We then describe the framework used to perform the studies described in this work. In Section \ref{sec:rec_hits} we describe the model to predict the number of photons arriving at a detector given only the position and scale of the energy deposition for both direct light as well as light reflected off the cathode of the TPC. In Section \ref{sec:parametrization} we describe the model to predict the distribution of arrival times of the photons for both of the above cases. We then test the performance of these models compared to a standalone Geant4 simulation, and to predictions obtained using lookup libraries. Finally, we show an example of the application of this model to a realistic event.

\section{Scintillation Light in Liquid Argon} \label{sec:description}

\subsection{Production of Scintillation Light in LAr}\label{sec:production}

The liquid argon scintillation light originates from the de-excitation of argon dimer states formed when an argon atom, excited or ionized by an interaction of a charged particle, attaches a neutral argon atom. This results in a relatively narrow emission wavelength with a peak of 128\,nm \cite{Doke:1988rq}, in the vacuum ultra-violet (VUV) range.
This mechanism is very prolific resulting in over 40000 photons being emitted per MeV of deposited energy in the absence of an electric field \cite{Doke:1990rza}. Applying an electric field reduces the recombination of argon dimers, decreasing the scintillation yield. At an electric field of 500\,V/cm, a typical value used in LArTPCs, the scintillation yield decreases to about 20000 photons/MeV \cite{Doke:2002oab}.

The time distribution of the emitted photons is composed of two exponential decaying functions with lifetimes of $\sim$6\,ns and $\sim$1.5\,$\mu$s, corresponding to the two possible dimer molecular states: the singlet and the triplet \cite{Hitachi:1983zz}. 

The amount of light emitted can decrease through quenching effects (Q) or recombination (R). Quenching is either caused by the ionisation density \cite{Cao:2014gns, Agnes:2018mvl} or because of contaminants \cite{Acciarri:2008kv,Acciarri:2008kx} that can absorb the energy of the de-excited dimer without emitting light. Recombination depends on the value of the electric field $\mathscr{E}$ in the argon as well as on the ionisation density.

The ratio between the amount of light emitted by each of the two components depends on the ionisation density created by the interacting particle, and is therefore used as a method of particle identification, e.g. in Dark Matter experiments \cite{Boulay:2006mb}. The decay times, particularly of the slow component, can also be affected by contaminants present in the argon such as nitrogen \cite{Acciarri:2008kv} and oxygen \cite{Acciarri:2008kx} through a quenching process where some dimers transfer their energy to the contaminants instead of de-exciting. The time composition can also be altered by doping with other noble gases, for example xenon \cite{Wahl:2014vma}.

\subsection{Transport of Scintillation Light in LAr}\label{sec:RS}

The scintillation light emitted by the argon dimers is able to travel long distances in the liquid argon. Its mean free path is primarily affected by Rayleigh scattering and absorption on contaminants. Rayleigh scattering does not change the number of travelling photons but deflects them on their path. This can be either detrimental or beneficial for the probability of light arriving at photon-detectors, depending on the position in the detector and distance travelled. Light undergoing scattering and still arriving at a photon-detector will have travelled a longer path than light impinging on the detector without any interactions. This leads to a non-trivial distribution of arrival times of the photons that could be interpreted as an apparent lengthening particularly of the fast component of the scintillation light. The value of the Rayleigh scattering length ($\lambda_{RS}$) is currently under intense study with different measurements and theoretical predictions reporting values from around 50\,cm \cite{Grace:2015yta, Calvo:2016nwp, Agnes:2017grb} all the way up to 110\,cm \cite{Neumeier:2015nha}. In this paper we use a value of 100\,cm which is close to the most recent reported value \cite{Babicz:2020den}, and inside of the range of the other expected values.

Argon scintillation photons can also be absorbed ($Q_{abs}$) by contaminants with a high cross-section for VUV photons such as nitrogen \cite{Jones:2013bca} and other elements that have been observed in commercial argon \cite{Calvo:2016nwp}. The total absorption can be modelled as an exponential suppression of the number of photons as a function of the distance travelled with the absorption length as a parameter.

\begin{figure}
\centering
\includegraphics[width=.5\textwidth]{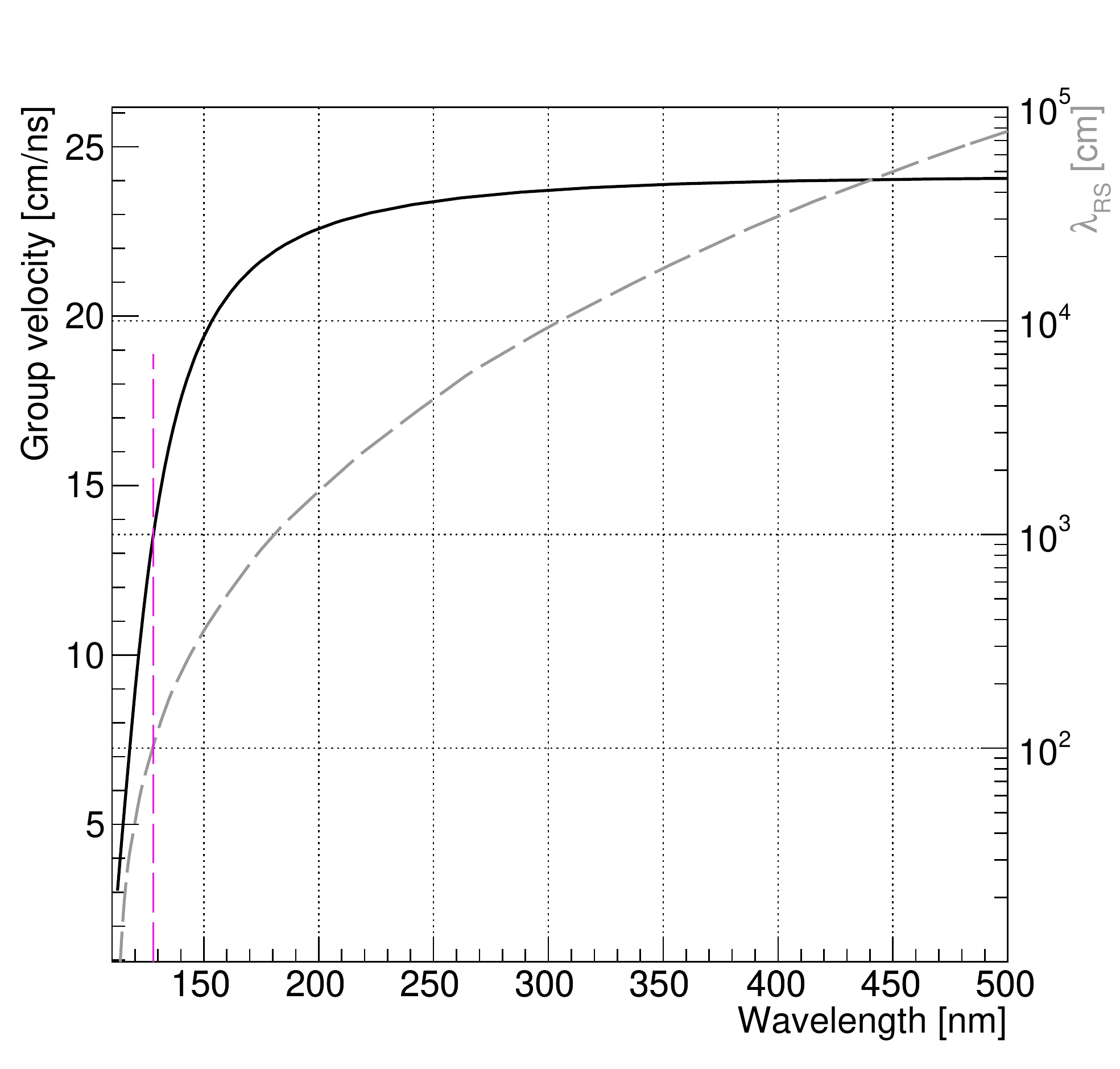}
\caption{Group velocity (left y-axis, solid line) and Rayleigh scattering length, $\lambda_{RS}$, (right y-axis, dashed line) as a function of the wavelength of the photons in liquid argon. Both spectra have been calculated using the constraint added by the group velocity measurement in~\cite{Babicz:2020den}. A line at 128\,nm is drawn to guide the eye to the scintillation emission wavelength in argon.} \label{fig:op_prop}
\end{figure}

Due to the high refractive index of liquid argon at VUV wavelengths \cite{Antonello:2004sx} the group velocity of photons emitted at 128\,nm is about two times slower than that of light at visible wavelengths, see Fig. \ref{fig:op_prop}.

\subsection{Detection of Scintillation Light in LAr} \label{sec:detection}

Detecting argon VUV scintillation light most often requires photon-detectors (PD) able to operate at, or close to, liquid argon temperatures. Cryogenic Photomultiplier tubes have been the default technology used in liquid argon detectors for some time \cite{Ankowski:2006xx,Nikkel_2007,Acciarri:2011qx} and have reported Quantum Efficiencies (QE) of up to 30\%. Recently, the idea of using Silicon Photomulipltiers (SiPM) has been gaining ground due to their low power consumption, small size, excellent noise performance at liquid argon temperatures and high QE up to 40\% \cite{Gola:2019idb}. SiPMs can be used as-is or enhanced using light-collectors such as light-guide bars \cite{Howard:2017dqb} or a light-trap, such as the ARAPUCA or X-ARAPUCA devices \cite{Segreto:2020jpd}. Another crucial challenge of detecting LAr scintillation light is the registration of VUV light before it is absorbed by the materials, e.g. glass or plastic, used to shield the sensitive area of the PD. The solution most commonly employed is to coat the PDs with a wavelength-shifting (WLS) compound, which absorbs the VUV light and emits light in the visible spectrum easily detectable by the PDs. The direction of the re-emitted light is random, so WLS-coated PDs suffer a $\sim$50\% decrease in efficiency due to light emitted away from the active surface. 

The travel time, $t_t$, of the scintillation light in large liquid argon detectors ranges from a few to several tens of nanoseconds. Wavelength-shifting compounds tend to impact the photon arrival time, because the emission of the visible light is not instantaneous and has an intrinsic decay time, $t_{WLS}$, which could delay the detection of the photons. 
The electronics and data acquisition chains in LArTPC detectors are usually designed with a resolution in a similar range, from 1-2\,ns to 16\,ns in most recent liquid argon neutrino experiments \cite{Acciarri:2016smi,Abi:2020evt,Ali-Mohammadzadeh:2020fbd}.  

In large LArTPCs used for neutrino experiments the PDs are usually placed behind planes of sense wires \cite{Acciarri:2016smi, Amerio:2004ze, Acciarri:2019wgd}. These and other components of the detector can introduce a further decrease in the number of detected photons due to shadowing effects ($Q_{trans}$). 

\subsection{Passive elements of detection, i.e. reflective foils}

The majority of materials used to construct LArTPC detectors absorb VUV photons causing them to be lost. A method to recover them used primarily in direct dark matter search detectors is to cover the walls of the detector with a highly reflective surface, e.g. ESR foils~\cite{Benetti:2007cd} or PTFE~\cite{Agnes:2014bvk}, coated with a wavelength-shifting material. These surfaces become passive elements of the light detection system and enhance the amount of light detected by the PDs. It should be noted that the light arriving at the PDs is now already shifted to visible wavelengths where the efficiency of the WLS-coated PDs could be different. Additionally, the group velocity and Rayleigh scattering length of photons at visible wavelengths in LAr are significantly different, see Fig. \ref{fig:op_prop}, which will have an impact in any transport effects.

Recently, neutrino experiments have begun exploring the method of installing WLS-coated reflector foils on the detector cathode. Examples include the LArIAT experiment (fieldcage and cathode) \cite{Acciarri:2019wgd} and SBND \cite{Basque2019_foils}. Implementing this solution has also been proposed for the DUNE detectors~\cite{Abi2020_tdr_iv}.

\subsection{Generic Model for Predicting Behavior of Scintillation Light Photons } \label{sec:genericmodel}

In general, the number of photons detected by a given PD from an energy deposit $\Delta E$ at position $(d,\theta)$ can be calculated using the formula:   
\begin{linenomath}
\begin{align}
\begin{split}
D_{\gamma} =& \Delta E \times  S_{\gamma}(\mathscr{E}) \times Q \times Q_{abs}(d) \times Q_{det} \times  \\ & Q_{trans}(\theta)  
 \times P(d,\theta) \times T(d,\theta), 
\end{split} \label{eq:first} 
\end{align}
\end{linenomath}
where the scintillation yield $S_{\gamma}(\mathscr{E})=R(\mathscr{E})/W_{ph}$ is the number of photons emitted per unit of deposited energy at an electric field $\mathscr{E}$. This is defined in terms of the work function, $W_{ph}$~=~19.5\,eV~\cite{Doke:1990rza}, or average energy needed to create a photon at $\mathscr{E} = 0$, and the recombination factor, $R(\mathscr{E})$, that accounts for the reduction of the scintillation yield due to the presence of an electric field. The position $(d, \theta)$ is defined in terms of the distance, $d$, between the energy deposit and the PD, and the {\it offset angle}, $\theta$, between the energy deposit and the normal to the PD surface. $Q$ is the quenching at emission, $Q_{det}$ is the PD efficiency, $Q_{abs}(d)$ is the loss due to absorption effects and depends on  $d$, and $Q_{trans}(\theta)$ represents the loss of transmission due to shadowing effects and depends on $\theta$. All of the $Q_x$ parameters have values in the range from 0 to 1. $P(d,\theta)$ is the geometric coverage of the PD and $T(d,\theta)$ signifies other transport effects, both of which depend on distance and angle. Most of the parameters in Eq.~\ref{eq:first} are independent of each other and during simulation can be applied at the stage that the light is generated, with the exception of $P$ and $T$ that are tied together and more complicated in their application. The focus of this paper is a method to estimate these two quantities.

Similarly, the time at which a photon is detected by a particular PD can be obtained by summing together the independent components resulting from the different processes that photons undergo: 
\begin{linenomath}
\begin{equation} \label{eq:tiempos}
t_{\gamma} = t_E + t_t(d,\theta) + t_{WLS} + t_{det},
\end{equation}
\end{linenomath}
where $t_E$ is the emission time determined from a distribution combining the dimer decay times with any quenching of the time components. $t_t(d,\theta)$ is the transport time resulting from the different paths the photons can take to arrive at the PD. $t_{WLS}$ is the time resulting from the intrinsic relaxation time of the wavelength-shifting compound. Finally, $t_{det}$ is the time due to the PD and electronics response. These components can be applied separately and independently. In this paper we also propose a model to calculate $t_t(d,\theta)$. 

\section{Simulation Framework}
\label{sec:simulation_framework}

To develop, test and validate the model of scintillation light transport described in this work, we compared it against a Geant4~\cite{linkgeant4} simulation embedded in the LArSoft software framework~\cite{Snider:2017wjd}.
Geant4 is capable of simulating liquid argon scintillation light emission, transport and boundary properties. We use these results as a baseline (i.e. our true information). In a Geant4 simulation, the user needs to provide the optical properties of the active medium, liquid argon, and all of the surrounding materials with which the photons can interact. The optical properties of LAr that we implemented are summarized in Table~\ref{tab:opprop}. Additionally, to account for potential contaminants in the detector we apply an absorption length of $\lambda_{abs}=20$\,m corresponding to 3\,ppm of nitrogen~\cite{Jones:2013bca}.

\begin{table}[h]
    \centering
    \begin{tabular}{|c|c|c|}
         \hline
         Parameter   & Type &  Value\\ \hline
         emission wavelength & spectrum & $\langle$128\,nm$\rangle$~\cite{Gurtler:1989em} \\
         fast component decay time  & number &  6\,ns~\cite{Segreto:2016jpl}  \\
         slow component decay time  & number &  1590\,ns~\cite{Segreto:2016jpl} \\
         refractive index & spectrum & $\langle$1.32$\rangle$~\cite{Babicz:2020den} \\
         Rayleigh scattering length & spectrum & $\langle$100\,cm$\rangle$~\cite{Babicz:2020den} \\
         \hline
        \end{tabular}
    \caption{Liquid argon properties used in the simulation.}
    \label{tab:opprop}
\end{table}

\begin{figure}
\centering
\includegraphics[width=.45\textwidth]{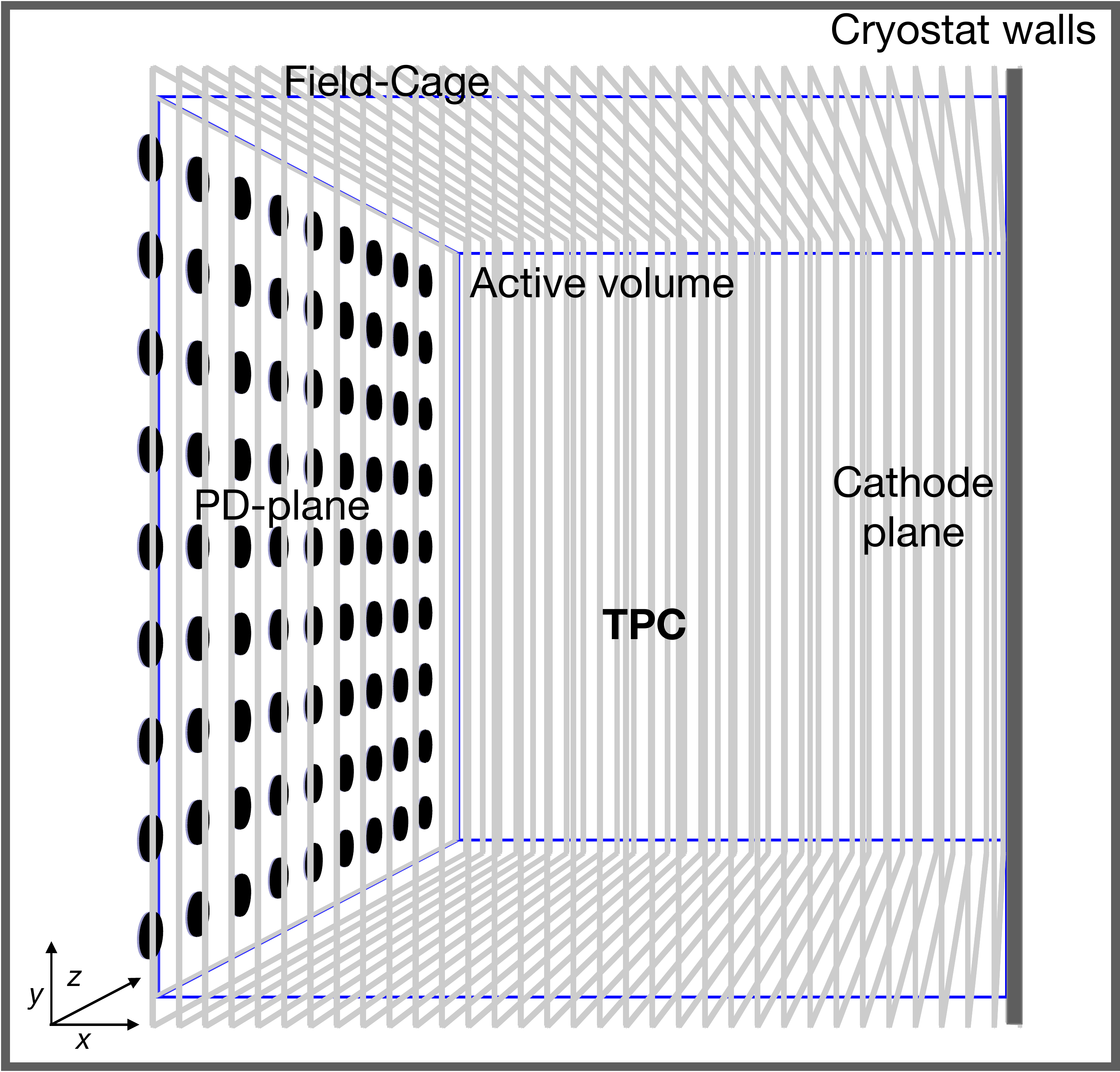}
\caption{Cartoon of the LArTPC detector geometry used in our simulation.} \label{fig:sbn-like-geometry}
\end{figure}

We define our simulated detector geometries using the GDML markup language~\cite{gdml:2006} to be realistic models of real-life LArTPC detectors. Figure \ref{fig:sbn-like-geometry} shows a schematic representation of our geometry. The main volume of liquid argon is delimited by metal walls of a cryostat\footnote{Photons leaving the field-cage are simulated and can in principle reflect off the cryostat walls and be detected on PDs.}. The field-cage surrounding the active volume is modelled as an array of metal strips on the top, bottom, upstream and downstream walls and spaced to provide $\sim$30\% optical transparency (a typical value in real detectors). The PDs are uniformly distributed in the open plane of the field-cage (left), referred to as the {\it photon-detector plane} (PD-plane). We simulate the PD sensitive windows as flat disks or rectangles facing the active volume. The cathode plane (right) is modelled as an opaque volume covered by polymeric reflector foils coated in a WLS. The WLS used in our simulations is TPB, for which the absorption/emission spectra are taken from~\cite{Francini:2013noa}. 

The reflectivities of the materials to VUV photons (pure scintillation emission) and visible photons (coming from the re-emission of the VUV photons absorbed by the TPB on the reflector foils) are listed in Table \ref{tab:matprop}.  In the simulations used in this work the reflections on optical boundaries have been modelled as Lambertian on a rough surface\footnote{In Geant4: GLISUR with polish = 0.5 for the argon-metal boundaries and GroundFrontPainted for WLS-foil boundary.} and we assume 100\% of light incident on the WLS is converted to visible wavelengths. 

\begin{table}[h]
    \centering
    \begin{tabular}{|c|c|c|}
         \hline
         Material & VUV reflectivity & Visible reflectivity \\ \hline
         metal &  25\% \cite{Zwinkels:94} & 60\% \cite{Zwinkels:94} \\
         reflector foils & N/A &  93\% \cite{Francini:2013noa}  \\
         \hline
        \end{tabular}
    \caption{Material reflectivities used in the simulation. Note that in our simulation the VUV reflectivity of the reflector foils is not defined as we assume 100\% of the incident light on the foils is converted to visible wavelengths.}
    \label{tab:matprop}
\end{table}
  
We test our model in two different geometries, corresponding to two different experiments employing LArTPC detectors: {\it subset-of-DUNE-like} and {\it SBND-like}. The main parameters describing these two geometries are listed in Table~\ref{tab:opdets}. 
In both geometries the PDs (PMT-like for the SBND-like geometry and X-ARAPUCA-like for the DUNE-like geometry) are distributed approximately evenly and with a PD located in the exact center of the PD-plane. 

\begin{table}[h]
    \centering
    \begin{tabular}{|c|c|c|}
    \hline
    Parameter & SBND-like & DUNE-like  \\ \hline
    width [cm]     & 200  & 365   \\
    height [cm]     & 400  & 1200  \\
    length [cm]     & 500  &  1400 \\
    number of PDs  & 66   & 123  \\
    PD shape     &  disk  &  rectangle   \\
    PD size     &  8''\,diameter &  $9.3\times9.3\,$cm$^{2}$ \\
    \hline
    \end{tabular}
    \caption{Summary of the main parameters of the two geometry definitions used.}
    \label{tab:opdets}
\end{table}

\begin{figure}
\centering
\includegraphics[width=.4\textwidth]{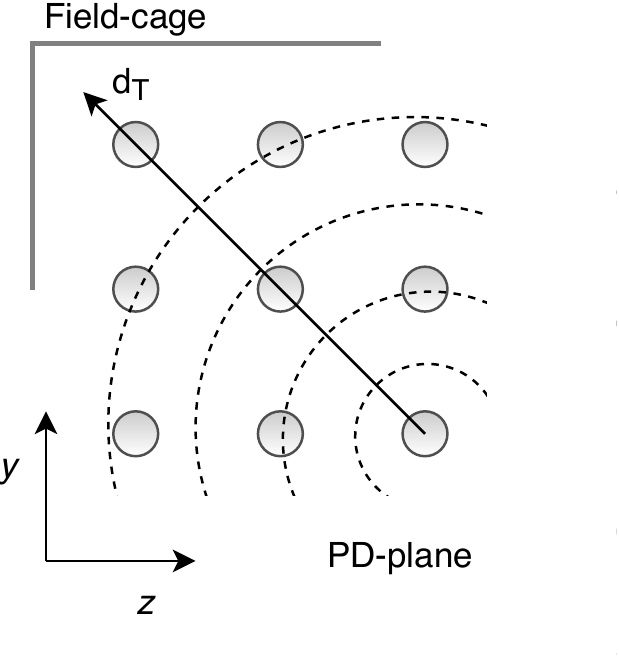}
\caption{Cartoon of the concentric cylinders at different radial distances, $d_{T}$, from the center of the photon-detector plane (PD-plane). Each grey disk represents a PD. Sets of energy depositions are simulated in each of these cylinders to fully cover the possible positions within the detector. } \label{fig:point_selection}
\end{figure}

We simulate energy depositions at points within the active volume that cover the full phase space of distances and angles between the scintillation and the PDs. To study border effects and cover different regions of the detector, we divide the volume into concentric cylinders at different radial distances, $d_{T}$, starting from the PD at the center of the PD-plane ($Y-Z$) outwards towards the corners of the field-cage, as illustrated in Fig. \ref{fig:point_selection}. Then at each $d_T$, energy depositions are simulated at evenly spaced positions in the drift direction, $X$, covering the full drift length. In the SBND-like geometry, the simulated energy depositions are spaced in approximately 20\,cm steps in the drift direction and 50\,cm steps in $d_T$. In the DUNE-like geometry, energy depositions are spaced in approximately 25\,cm steps in the drift direction and 100\,cm steps in $d_T$. The step sizes in $d_T$ are driven by the distances between the PDs, while those in the drift direction have been roughly chosen to cover the entire detector volumes without requiring an excessive total number of points. Five different energy deposition locations are simulated for each $X$ and $d_T$ pair, which define a cylindrical shell in three dimensions. The first of these is chosen to be directly in front of a PD and the remaining four placed at increasing offsets of approximately 5\,cm in $d_T$ to ensure a sufficiently large population of photons incident on the PDs from small angles. Following the above method, we have simulated approximately 2500 and 3500 energy deposition points in the SBND-like and DUNE-like geometries respectively.

Energy depositions generating $25\times10^6$ photons ($\sim$1\,GeV) are simulated in the SBND-like geometry and $100\times10^6$ photons ($\sim$4.2\,GeV) in the DUNE-like geometry. The simulation takes approximately between 90\,s and 170\,s per $1\times10^6$ photons generated, depending on the position in the detector. The larger number of photons in the DUNE-like case was chosen to improve the statistics in the number of photons incident on the PDs from the larger possible distances in this geometry. 

\section{Predicting the Number of Detected Photons} \label{sec:rec_hits}

In this section we develop a model capable of predicting the number of photons arriving at any given photon detector, based on the size of an energy deposit and the position where it occurred.  
We first focus on the VUV light that travels to the PDs directly from the point of emission, which we call the {\it direct} component. 
Starting from basic geometric considerations, we first calculate the solid angle subtended by a PD in an ideal infinite detector. We then account for the presence of Rayleigh scattering in the propagation by implementing a correction to the amount of light predicted geometrically. Finally, we add an extra correction to account for border effects present in real detectors of finite size. 

The model developed to predict the number of VUV photons is then used as the first stage of a second model that predicts the number of photons arriving at PDs after being converted to visible wavelengths and reflected by a WLS-coated reflective detector cathode. We call this the {\it reflected} component of scintillation light. This model also employs corrections for transport effects and takes into account the finite size of the detector. 

\subsection{Direct VUV Light}\label{sec:vuv_rec}
\subsubsection{Geometric Considerations}
Scintillation light is emitted isotropically. This means that in an ideal case the number of photons arriving at a given PD could be calculated by simply estimating its geometric acceptance with respect to the scintillation point, i.e. the solid angle subtended by the sensitive window. This is a calculation that can be performed either analytically or with simple numerical integration depending on the shape of the PD. Such solutions exist, e.g. for disk~\cite{Paxton:1959} and rectangular~\cite{Khadjavi:1968} shapes, that cover most existing PD designs. The conclusions drawn in this work can be extrapolated to any other PD shape, once the calculation of the solid angle subtended from a point-like source by such a shape is known.

\begin{figure}
\centering
\includegraphics[width=.5\textwidth]{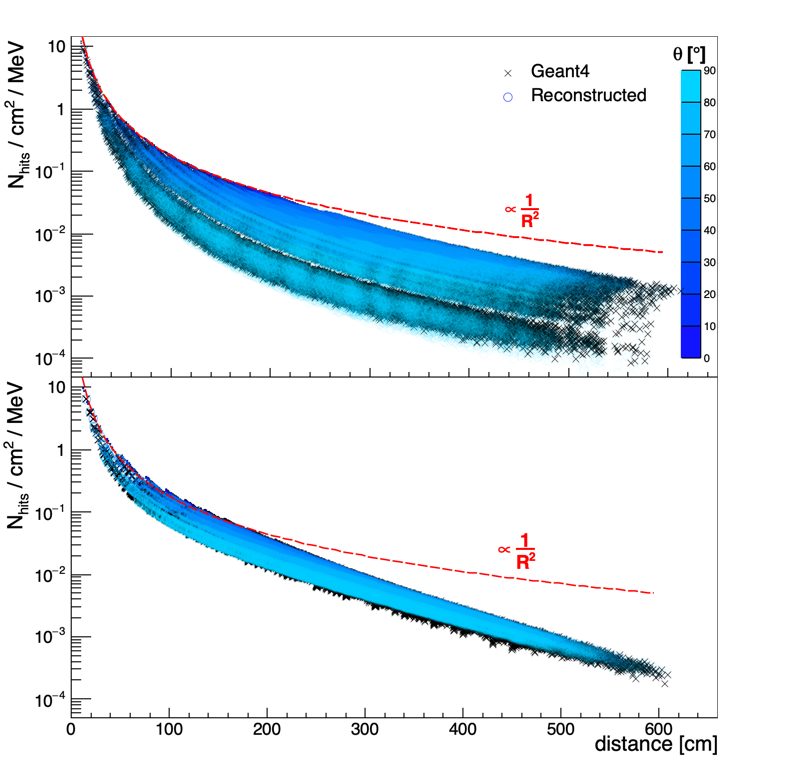}
\caption{Top: Number of Geant4 tracked (black crosses) and analytically predicted (blue circles) scintillation photons arriving at the PDs per unit of deposited energy and PD sensitive-window area, in the SBND-like detector geometry. In this case Rayleigh scattering is not included and all material reflectivities are set to zero. The red dashed line represents a pure 1/$R^2$ behavior. It diverges from the simulated points when the size of the detector excludes any further points on-axis to the PDs, at $d = 200$\,cm. Note that the dependence on the offset-angle $\theta$ is indicated by the shades of blue of the reconstructed circle-markers. Bottom: Variation of the top scenario where Rayleigh scattering~\cite{Babicz:2020den} is included. Rayleigh scattering strongly shapes the amount of light observed in the PDs.} \label{fig:recHits_scatter}
\end{figure}

This approach works in an idealized case: in the absence of Rayleigh scattering and reflections by the detector materials (i.e. $\lambda_{RS}\to\infty$ and all materials 100\% absorptive). In this scenario, the calculation becomes purely geometric and for any given energy deposition, $\Delta E$, we can calculate the number of photons incident on a PD as,
\begin{linenomath}
\begin{equation} \label{eq:dNhits}
N_{\Omega} = e^{-\frac{d}{\lambda_{abs}}} \Delta E\cdot S_{\gamma}(\mathscr{E})\frac{\Omega}{4\pi},
\end{equation}
\end{linenomath}
where the $S_{\gamma}(\mathscr{E})$ is the scintillation yield of LAr for a given electric field, and $\Omega$ is the subtended solid angle. We also implement absorption effects due to contaminants: $Q_{abs} = e^{-\frac{d}{\lambda_{abs}}} $, where $\lambda_{abs}$ is the absorption length and $d$ is the distance to the PD. The performance of Eq.~\ref{eq:dNhits} at predicting the number of photons can be seen in the top panel of Fig.~\ref{fig:recHits_scatter}. This shows a comparison between the number of photons hitting the PD windows predicted by Eq.~\ref{eq:dNhits} and the number obtained from a full Geant4 simulation, normalized to the sensitive-window area and the energy deposited. The pure-geometric calculation agrees with the full simulation within expected Poisson fluctuations. The gradient-colors of the circles represent the offset angle, $\theta$ defined in Section~\ref{sec:genericmodel}. It can be seen that more light is observed by the PDs from emission points closer and more on-axis, as expected. The red dashed line indicates a perfect 1/$R^2$ behavior. Even in this simple case, it becomes clear that it is necessary to account for the offset angle as that can change the prediction for a given distance by up to two orders of magnitude. 

\begin{figure}
\centering
\includegraphics[width=.5\textwidth]{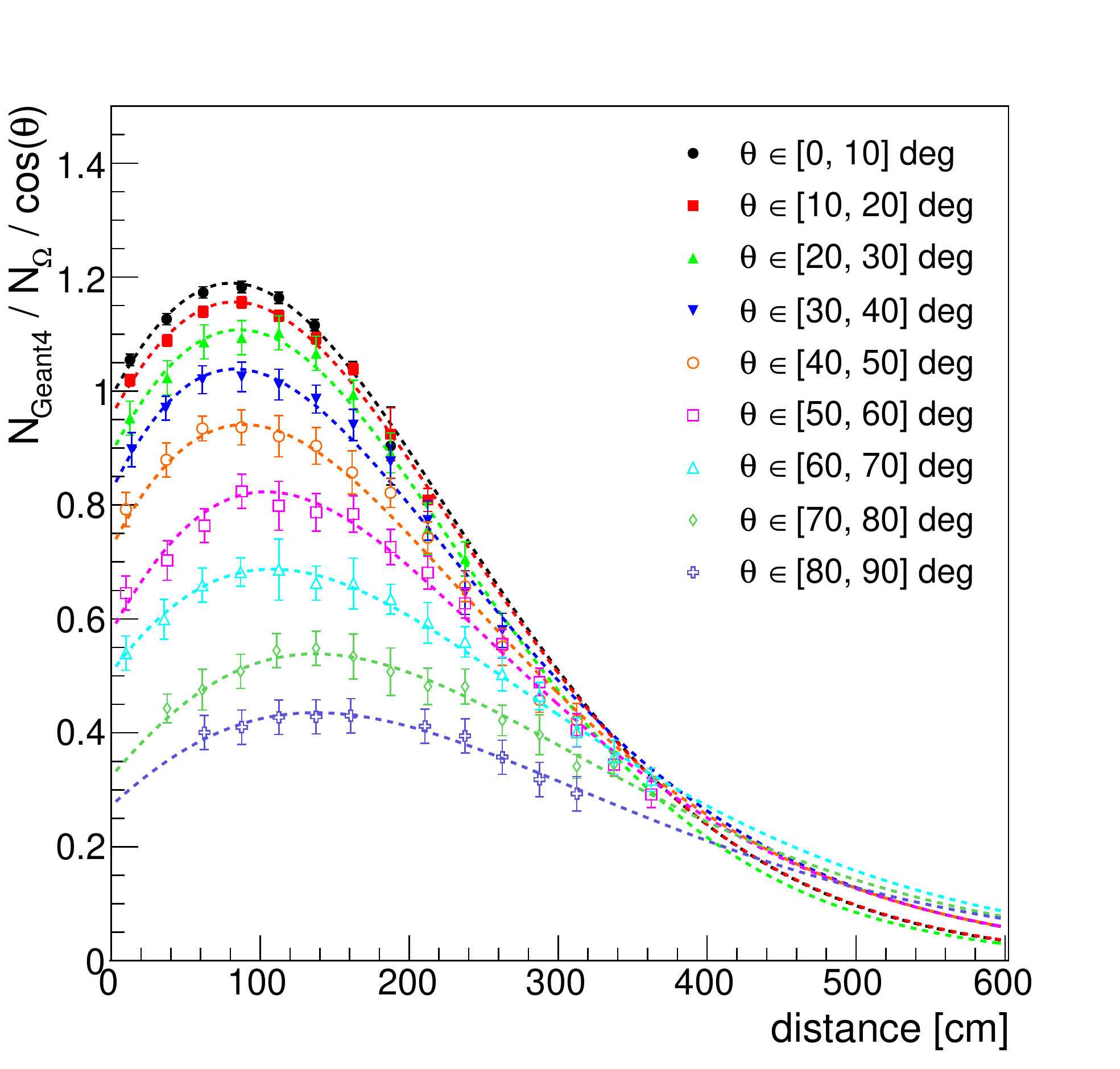}
\caption{Relation between the number of Geant4 simulated hits on the PDs and the pure geometric estimation described by Eq.~\ref{eq:dNhits}, in the SBND-like geometry. The error bars represent the standard deviation of the distribution within each angular bin. A strong dependency is clear in both distance and offset angle. At any angle, the dependency of the ratio with distance can be accurately described by a Gaisser-Hillas function, as illustrated by the dashed curves. To avoid large divergences at big offset angles we have found it more convenient to work with the projected solid angle.}\label{fig:SBN_rainbow}
\end{figure}

\subsubsection{Corrections to the Geometric Approach}
\label{sec:GHcorrections}
The basic solid angle approach breaks down when Rayleigh scattering is introduced into the simulation. The VUV scintillation photons in LAr undergo scattering during propagation with a characteristic length, $\lambda_{RS}$, that is small compared to the size of current and future LArTPC experiments. This alters the path of the majority of the photons, and consequently the number of them arriving at the PDs. Once Rayleigh scattering is included in the Geant4 simulation the distribution of points in Fig.~\ref{fig:recHits_scatter}-top is altered to that shown in the bottom panel. The Rayleigh scattering significantly alters the amount of light observed in the PDs, and it is therefore essential to account for its effect during simulation. These effects strongly depend on both the distance, $d$, and the offset angle, $\theta$, of the PD relative to the light emission point.

To build corrections for the effects of Rayleigh scattering, we calculate the ratio between the number of incident photons from Geant4 simulation, $N_{hits}$, and the geometric estimation from Eq.~\ref{eq:dNhits} projected on cos$(\theta)$, $N_{\Omega}$/cos($\theta$). For simplicity, we split the phase space into 10$^{\circ}$-wide bins in $\theta$. The discretization in $\theta$ introduces a systematic effect in our model: a more (less) sampled choice would result in a more (less) accurate correction. Our choice in this work is a trade off between accuracy and computational time. The resulting ratios, shown in Fig.~\ref{fig:SBN_rainbow}, are smooth distributions as a function of distance and clearly separated between the different angular bins. This indicates that a parameterization in $(d, \theta)$ should include all the main dependencies and consequently be sufficient to predict the number of arriving photons. 
We find that the distributions shown in Fig.~\ref{fig:SBN_rainbow} can, for all angles, be accurately described using Gaisser-Hillas (GH) functions~\cite{1977ICRC....8..353G}, as illustrated by the dashed curves:
\begin{linenomath}
\begin{equation} \label{eq:GH_basic}
GH(d) = N_\text{max}\left(\frac{d-d_0}{d_\text{max}-d_0}\right)^{\frac{d_\text{max}-d_{0}}{\Lambda}}e^{\frac{d_\text{max}-d}{\Lambda}},
\end{equation}
\end{linenomath}
where $N_{max}$ is the maximum of the function located at a distance $d_{max}$, and $d_0$ and $\Lambda$ are parameters describing the width of the distribution. We implement the GH functions as the core of our numerical model to predict the scintillation light signals in large LArTPC detectors: (i) the number of incident photons on each PD is predicted by the solid angle that the aperture of the detector subtends, (ii) then the effect of Rayleigh scattering is accounted for via corrections to the geometric prediction\footnote{As the Gaisser-Hillas function can be shown to be equivalent to a Gamma distribution, the latter could be used with similar results. For mathematical convenience, we have chosen to use the Gaisser-Hillas definition.}. Once we apply these corrections to Eq.~\ref{eq:dNhits}, our model precisely predicts the number of incident photons on each PD as shown qualitatively in Fig.~\ref{fig:recHits_scatter} (bottom panel, blue circles). A quantitative comparison is discussed in Section~\ref{sec:validation-vuv}.

In the limit d $\to$ 0, the effect of Rayleigh scattering should be negligible and the y-intercept in our corrections should correspond to the value cos($\theta$) (i.e. $\sim$1 for the on-axis case). The Gaisser-Hillas-like shape of the corrections suggests a behavior of the light such that for small distances from the PD, the probability of detecting scattered photons that would otherwise escape from the detectors is larger than the fraction that is lost due to the scattering. This situation continues for larger distances until a point at which it is reversed and more photons are lost than gained. 
Additionally, once taking into account the 1/cos(${\theta}$) factor in Fig. \ref{fig:SBN_rainbow}, we can see that PDs at large $\theta$ (that have a small geometric acceptance) present a higher relative probability to recover scattered photons compared with PDs located closer to on-axis.
This behavior explains the significant tightening of the angular dependence in the points on Fig.~\ref{fig:recHits_scatter} when Rayleigh scattering is included (bottom) compared to the ideal case (top), although the dependence remains strong. These effects also result in the detector size having an impact on the required correction curves:
the greater the active volume in which photons can scatter, the greater the probability that these photons will end up feeding the signal. We account for this effect next.
\subsubsection{Correcting for Detector Size: Border Effects}
\label{sec:BordersVUV}
The dependency of our derived corrections on the detector size can be treated as a border effect. These borders (i.e. the cryostat walls and other detector components) not only delimit the active volume where photons can travel and scatter, but also consist of surfaces that can reflect or absorb them. These effects influence the amount of light observed in the PDs and, as a consequence, different sets of corrections may be required for different regions of the liquid argon volume.

\begin{figure}
  \includegraphics[width=.48\textwidth,keepaspectratio]{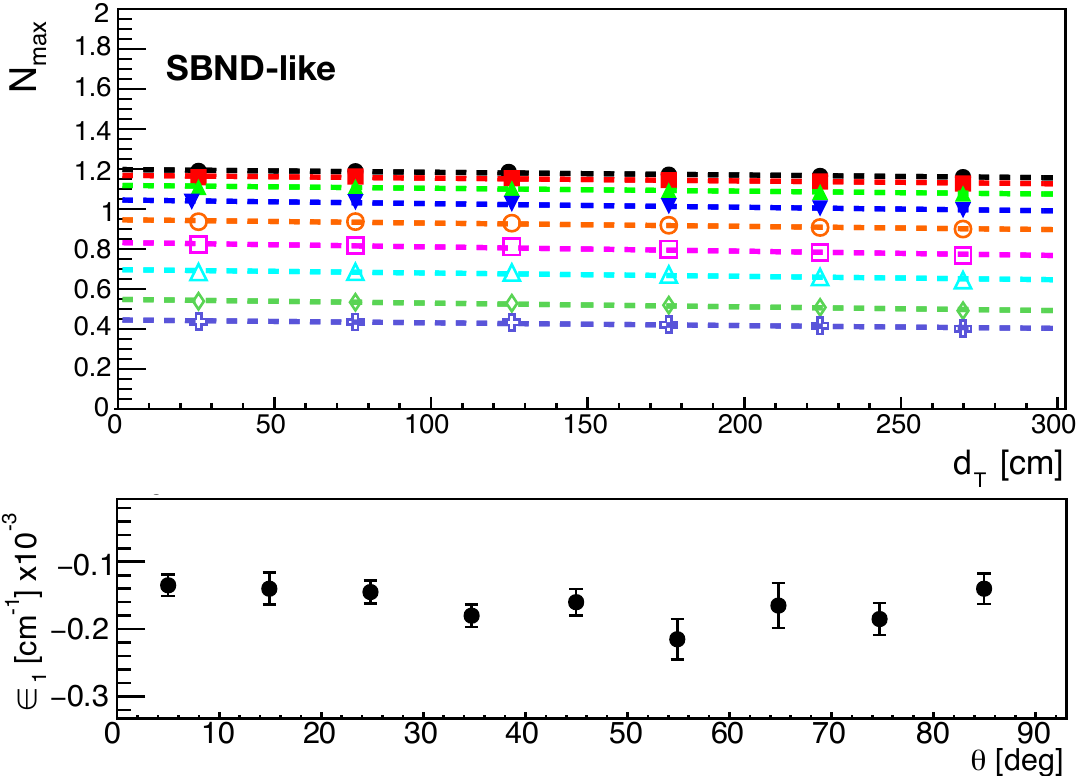}
  \includegraphics[width=.48\textwidth,keepaspectratio]{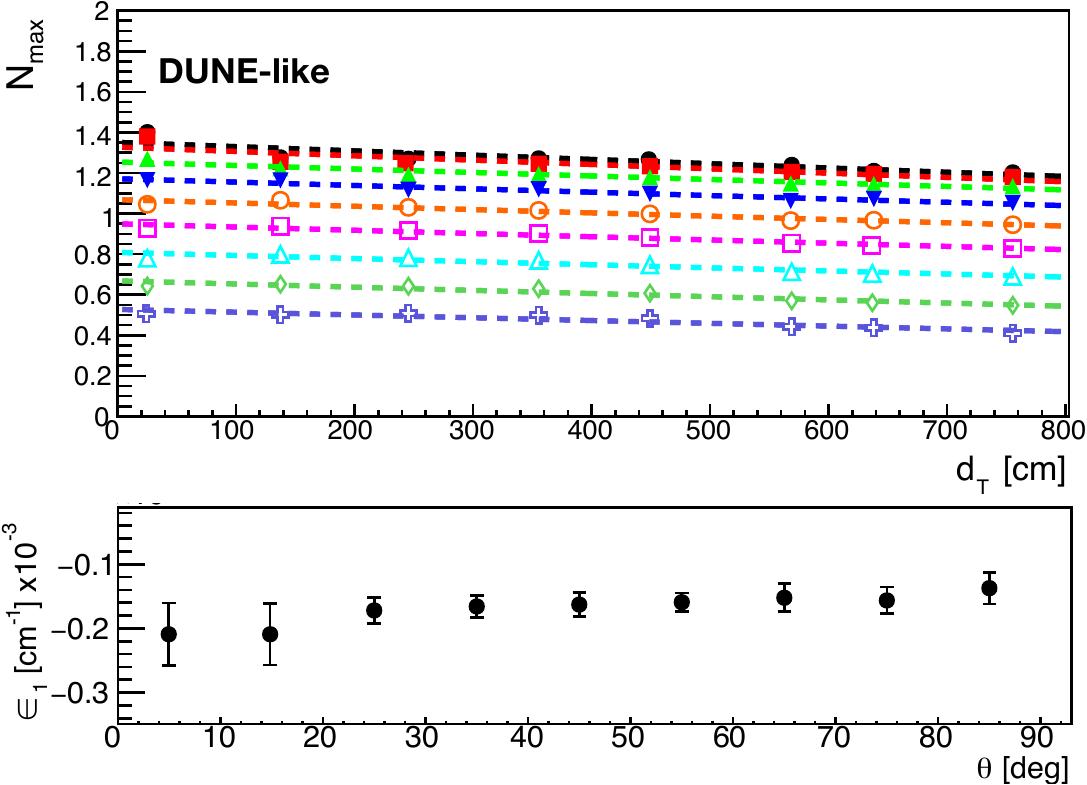}
  \caption{$N_{max}$ Gaisser-Hillas parameter dependency on distance to PD-plane center, $d_{T}$, for the SBND-like (top) and DUNE-like (bottom) geometries. The different colors refer to different $\theta$ bins as shown in Fig.~\ref{fig:SBN_rainbow}. The lines represent the linear fit of the points. The slopes, $\epsilon_{1}$, of the linear fits for the different offset angles are shown in the lower panels.}
  \label{fig:border_slopes1_both}
\end{figure}

\begin{figure}
\centering
\includegraphics[width=.5\textwidth]{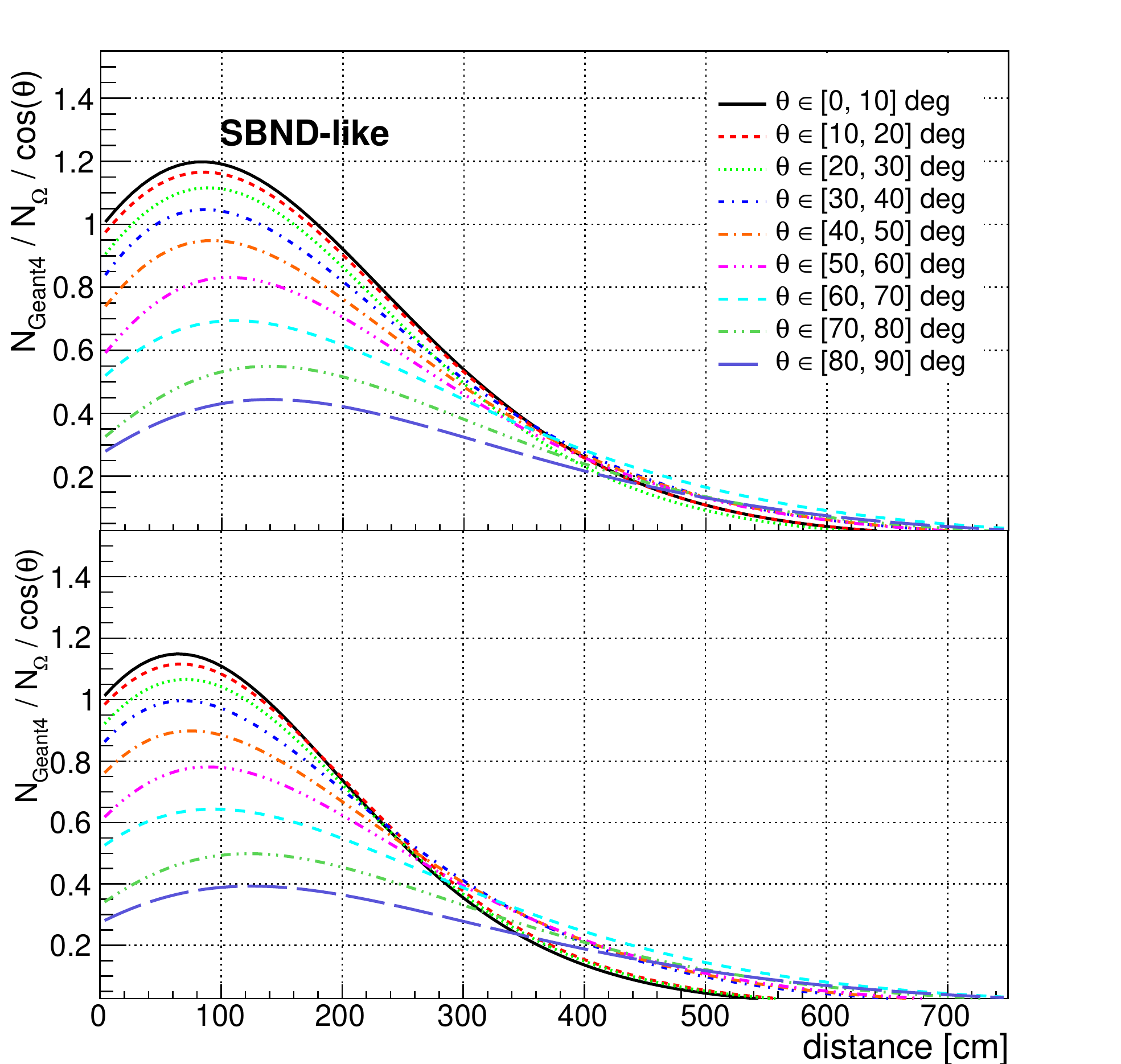}
\includegraphics[width=.5\textwidth]{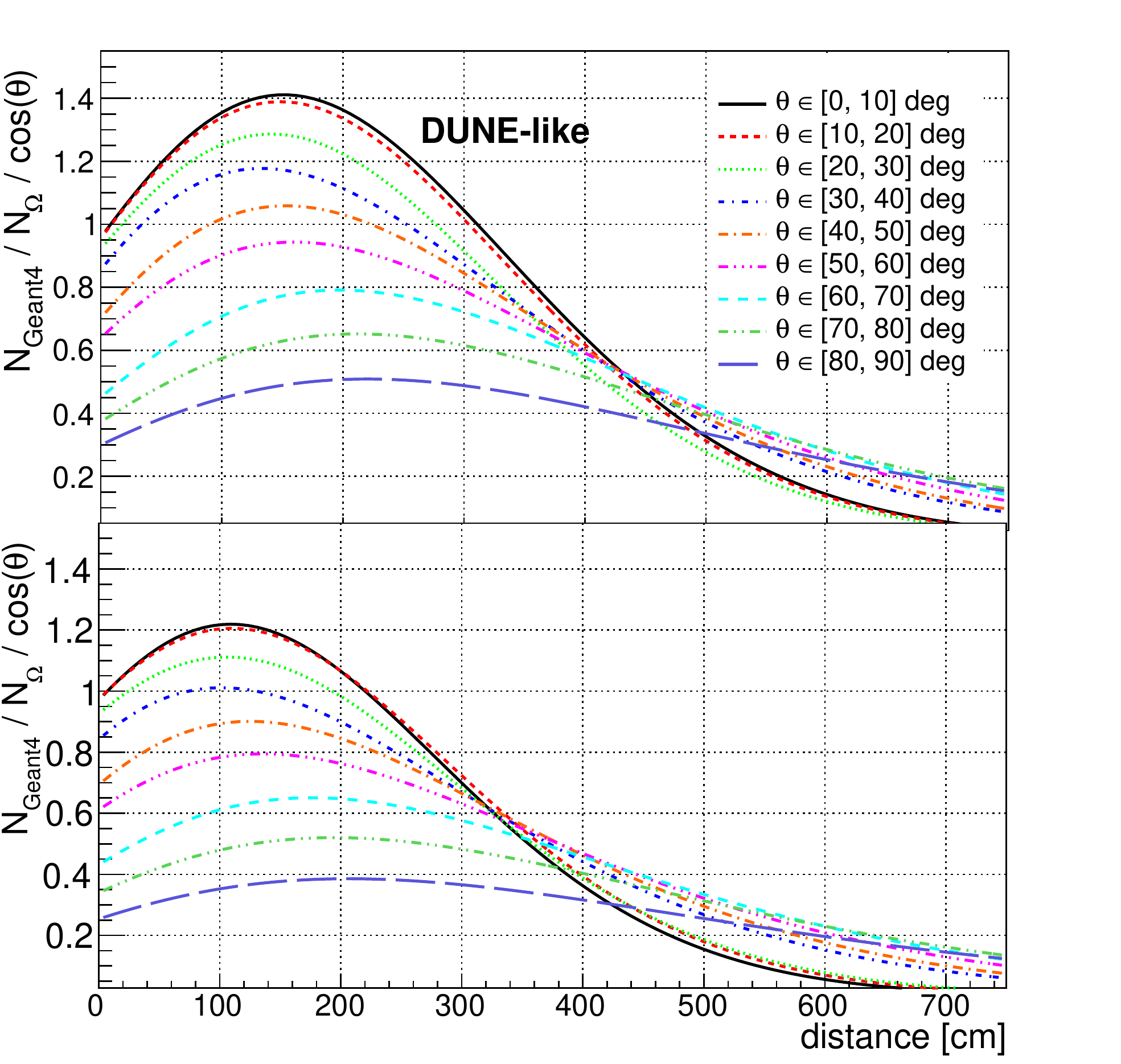}
\caption{Correction curves for the two geometries under study, for scintillation in the center of the TPC (top) and in the farthest corner (bottom).} 
\label{fig:border_corr_both}
\end{figure}

To develop the corrections, we examine the behavior of the parameters of the Gaisser-Hillas functions as a function of the radial distance $d_T$ as defined in Section~\ref{sec:simulation_framework}. 
For simplicity, and taking advantage of the strong correlation between the $d_0$ and $\Lambda$ parameters of the Gaisser-Hillas functions, we fix the value of $d_0$ absorbing all of the $d_T$ dependencies into the remaining three parameters. Figure \ref{fig:border_slopes1_both} shows the results for the $N_{max}$ parameter (similar results are obtained for $d_{max}$  and $\Lambda$, and are shown in \ref{sec:appendix_vuv}). We observe a linear dependency in $d_T$ for all of the offset angle bins in both geometries under study. There is also a weak dependency in the slopes of the lines with $\theta$, increasing for the more off-axis angles. We take these dependencies into account to accurately estimate the number of scintillation photons arriving at a PD for the entire active LAr volume. To this aim, we redefine the Gaisser-Hillas parameters in Eq.~\ref{eq:GH_basic} as:
\begin{linenomath}
\begin{equation} \label{eq:GH_borders}
\begin{split}
N^\prime_\text{max} = N_\text{max} + \epsilon_1(\theta) d_{T} \\
d^\prime_\text{max} = d_\text{max} + \epsilon_2(\theta) d_{T} \\
\Lambda^\prime = \Lambda + \epsilon_3(\theta) d_{T},
\end{split}
\end{equation}
\end{linenomath}
where $N_\text{max}, d_\text{max}$ and $\Lambda$ are the values of the parameters in the center of the PD-plane ($d_{T}=0\,$cm), and $\epsilon_1, \epsilon_2$ and $\epsilon_3$ are the slopes of the linear corrections for each parameter respectively. This new function is referred to as $GH^\prime$. To give an indication how these corrections affect the probability of photons arriving at PDs, Fig.~\ref{fig:border_corr_both} shows examples of correction curves for the two extreme cases, center (top) versus corner (bottom), for both the SBND-like and DUNE-like geometries.

Bringing all of the above effects together, the model we describe here is able to estimate the number of detected scintillation light photons using  Eqs.~\ref{eq:dNhits},~\ref{eq:GH_basic}~and~\ref{eq:GH_borders}, combined as:
\begin{linenomath}
\begin{equation} \label{eq:Nvuv_rec}
N_{\gamma} = N_{\Omega} \times GH^{\prime}(d, \theta, d_{T}) / cos(\theta),
\end{equation}
\end{linenomath}
which depends only on the distance and angle between the emission point and the PD, and distance of the emission point from the center of the detector. Note that $N_{\gamma}= \Delta E \times S_{\gamma} \times Q_{abs} \times P(d,\theta)\times T(d,\theta)$ using the notation from Eq.~\ref{eq:first}.

\subsection{Reflected Visible Light}
\label{sec:NHits_vis}

\subsubsection{Basic Geometric Model}

The number of photons arriving at the PDs in a LArTPC that has wavelength-shifting reflector foils on the cathode can also be predicted using a geometric approach. This requires expanding the model developed for the direct light VUV-only case described in Section \ref{sec:vuv_rec}. The prediction of the wavelength-shifted and reflected {\it visible} light is inherently dependent on the specific detector geometry, because the distance between the reflective foils and the PDs becomes a key element of the model. Additionally, unlike for the VUV light, the wavelength-shifted light is much more likely to reflect from the borders of the detector and the field cage.
Therefore, we construct the model for wavelength-shifted light using a realistic detector geometry from the start rather than using an idealized detector. 

\begin{figure}
\centering
\includegraphics[width=.45\textwidth]{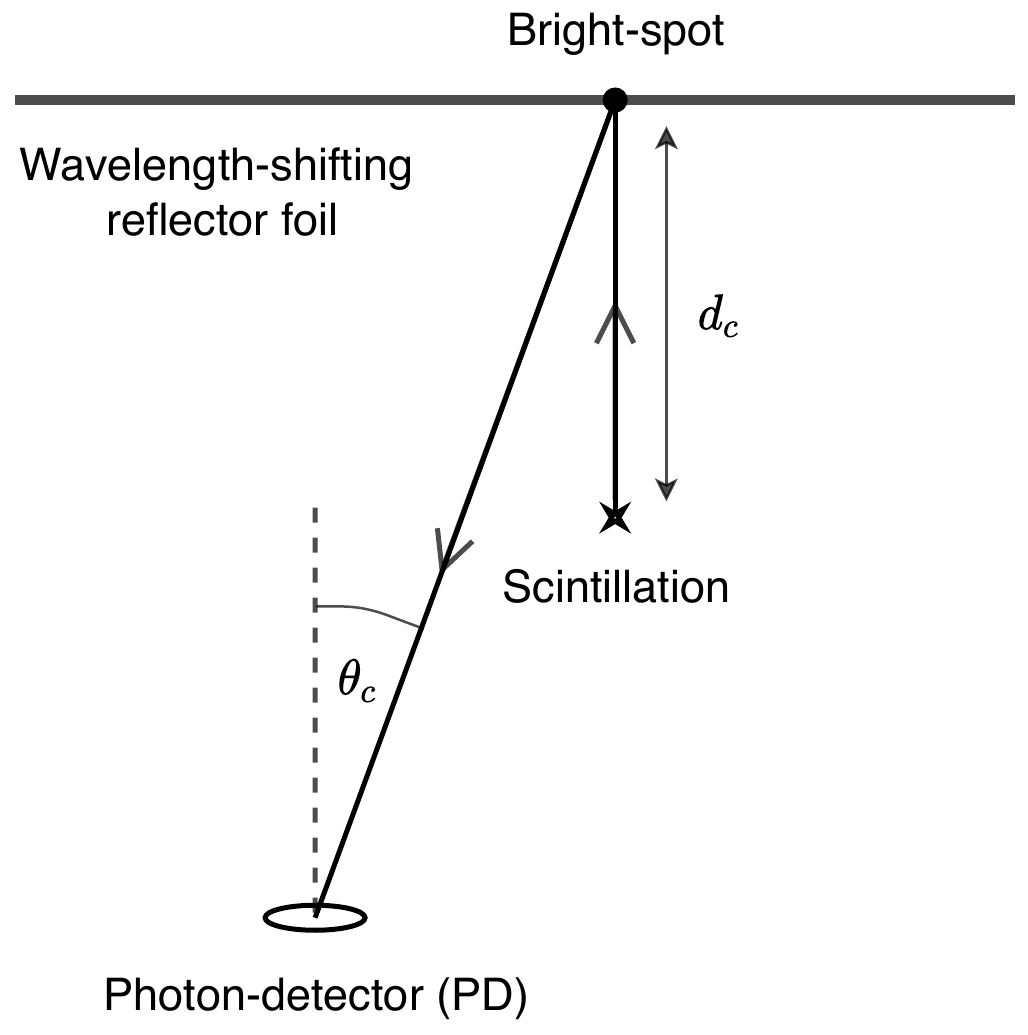}
\caption{Diagram illustrating the geometric model for predicting the number of photons incident on the PDs as a result of wavelength-shifting reflector foils on the detector cathode and predicting the arrival time distribution of these photons on the PDs.}\label{fig:vis_model_diagram}
\end{figure} 

The visible light arrives at the PDs after being re-emitted and possibly reflected by the WLS-coated reflector foils at the cathode of the detector. Therefore, we first calculate the number of VUV photons incident on the reflector foils using the solid angle that the entire cathode subtends, $\Omega_{c}$. This is corrected for the effects of Rayleigh scattering using Eq.~\ref{eq:Nvuv_rec}. We then assume that these photons are re-emitted approximately isotropically after being wavelength-shifted and reflected,
and that the region of the cathode in front of the scintillation in the drift direction will be the dominant source of the visible photons.
The central point of this region is referred to as the {\it bright-spot}, and is illustrated in Fig.~\ref{fig:vis_model_diagram} together with the other elements of the geometric model for the reflected light. The number of photons incident on each PD can then be calculated using the solid angle subtended by the PD aperture as viewed from the bright-spot, $\Omega_{PD}$. The geometric prediction for the number of visible photons arriving at the PDs can therefore be expressed as,
\begin{linenomath}
\begin{equation} \label{eq:Nhits_vis}
N_{\Omega, reflected} = N_{\gamma, direct}(\Omega_c, d_c, \theta_c, d_T) \times Q_r \times \frac{\Omega_{PD}}{2\pi},
\end{equation}
\end{linenomath}
where $N_{\gamma, direct}(\Omega_c, d_c, \theta_c, d_T)$ is the prediction of the number of photons incident on the cathode using the direct VUV light model given by Eq.~\ref{eq:Nvuv_rec} and $Q_r = Q_{WLS} \times Q_{foil}$ is a scaling factor accounting for the WLS efficiency, $Q_{WLS}$, and the foil reflectivity, $Q_{foil}$. The solid angle of the PD, $\Omega_{PD}$, is divided by $2\pi$ rather than $4\pi$ due to the presence of the highly reflective foils beneath the WLS.

\subsubsection{Corrections for PD Position}

The basic geometric model provides an initial approximation of the number of reflected photons incident on each PD.  
The assumption that the bright-spot region is dominant does not fully account for the distribution of the re-emitted wavelength-shifted photons across the whole surface of the reflective cathode. The approximation performs well for the PDs placed close to on-axis (at small $\theta_c$) that see the majority of the light. However, it is a poorer approximation for the PDs located further off-axis where a larger fraction of the observed light originates from regions of the cathode opposite to the PD rather than the bright-spot. We therefore implement corrective factors to the basic model to account for these effects in an analogous way to the direct light model described in Section \ref{sec:vuv_rec}. 
Because the corrections are developed in a realistic geometry, they also account for effects of reflections of the wavelength-shifted photons from the field-cage and the walls of the cryostat.

\begin{figure}
	\centering
	\includegraphics[width=.5\textwidth]{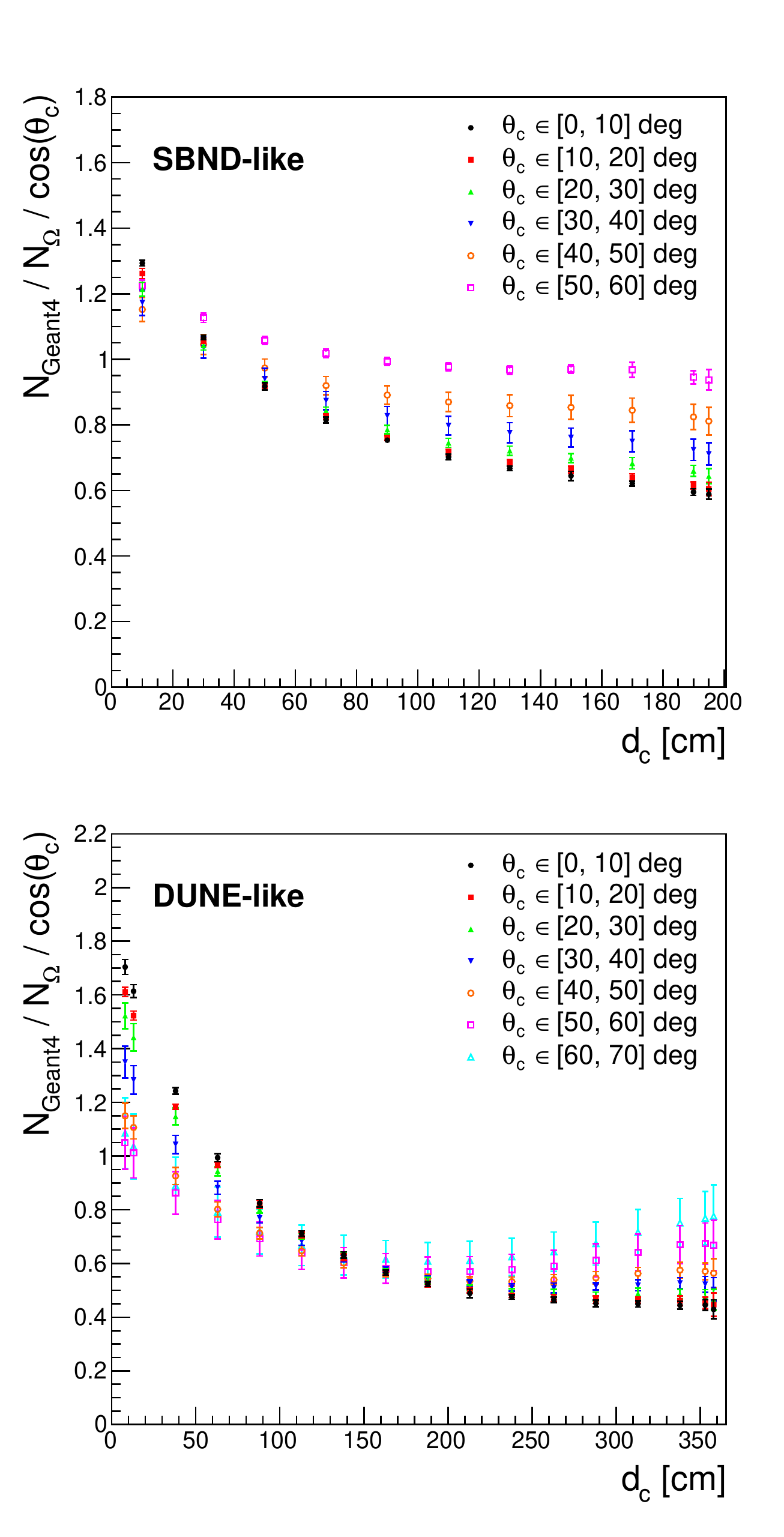}
	\caption{Ratio between the number of photons incident on the PDs in Geant4 simulation and the prediction from the reflected light geometric model for scintillation occurring in the central region of the SBND-like and DUNE-like detector geometries.}
	\label{fig:vis_centre_corrections}
\end{figure}

Similar to the method used for the direct light model, the required corrections are taken as the ratio between the number of incident photons on the PDs in Geant4 simulation and the prediction from the basic geometric model. For scintillation photons generated in the central region of a detector, the difference between the full Geant4 simulation and the model can be parameterized using only the distance between the scintillation and the bright-spot, $d_c$, and the offset angle between the bright spot and the normal to the PD surface, $\theta_c$. Examples of the parameterized corrections are shown in Fig.~\ref{fig:vis_centre_corrections} for the SBND-like and DUNE-like detector geometries. In both cases, the maximum offset angle as viewed from the bright-spot is defined by the size of the detector geometry. The error bars represent the standard deviation of the distribution within each angular bin. Its value can be affected significantly by the detector size, as can be seen comparing the DUNE-like case with the SBND-like one. This is caused by the large $\theta_c$ angular bins describing larger regions of the detector volume where effects arising from reflections on the detector walls can be significantly different. We note that the scintillation points with the highest variations, i.e. at high $d_c$ or $\theta_c$, account for a relatively small fraction of the total light observed. 

To calculate the PD-position corrective factors we employ the mean of the $N_{Geant4}/N_{\Omega}/cos(\theta_c)$ distributions within each angular bin. Instead of using a fit to the corrective factors, we use linear interpolation in $d_c$ to find the exact correction for the prediction from the geometric model. 

\subsubsection{Correcting for Scintillation Position:  Border Effects}

In addition to the corrective factors accounting for the position of the PDs with respect to the point where the scintillation light was emitted, further corrections are required to account for the position where the light is created inside of the detector. Scintillation light created closer to the walls will be significantly affected by their proximity. The wavelength-shifted photons can be reflected off the walls, while the VUV light can be absorbed before it reaches the WLS-coated cathode plane.

We again account for these effects using parameterized corrective factors. We find that, similar to the PD-position based corrections, they depend on $d_c$ and $\theta_c$. An additional parameter is the position of the scintillation emission relative to the borders of the detector volume. Therefore, similar to the direct light model border corrections described in section \ref{sec:BordersVUV}, sets of corrective factors are created at different distances, $d_T$, from the center of the detector, as illustrated in Fig. \ref{fig:point_selection}. Then, during simulation, linear interpolation is used in both $d_c$ and $d_T$ for the required angular bin in $\theta_c$ to calculate the exact corrective factor required.

\begin{figure}
	\centering
	\includegraphics[width=.5\textwidth]{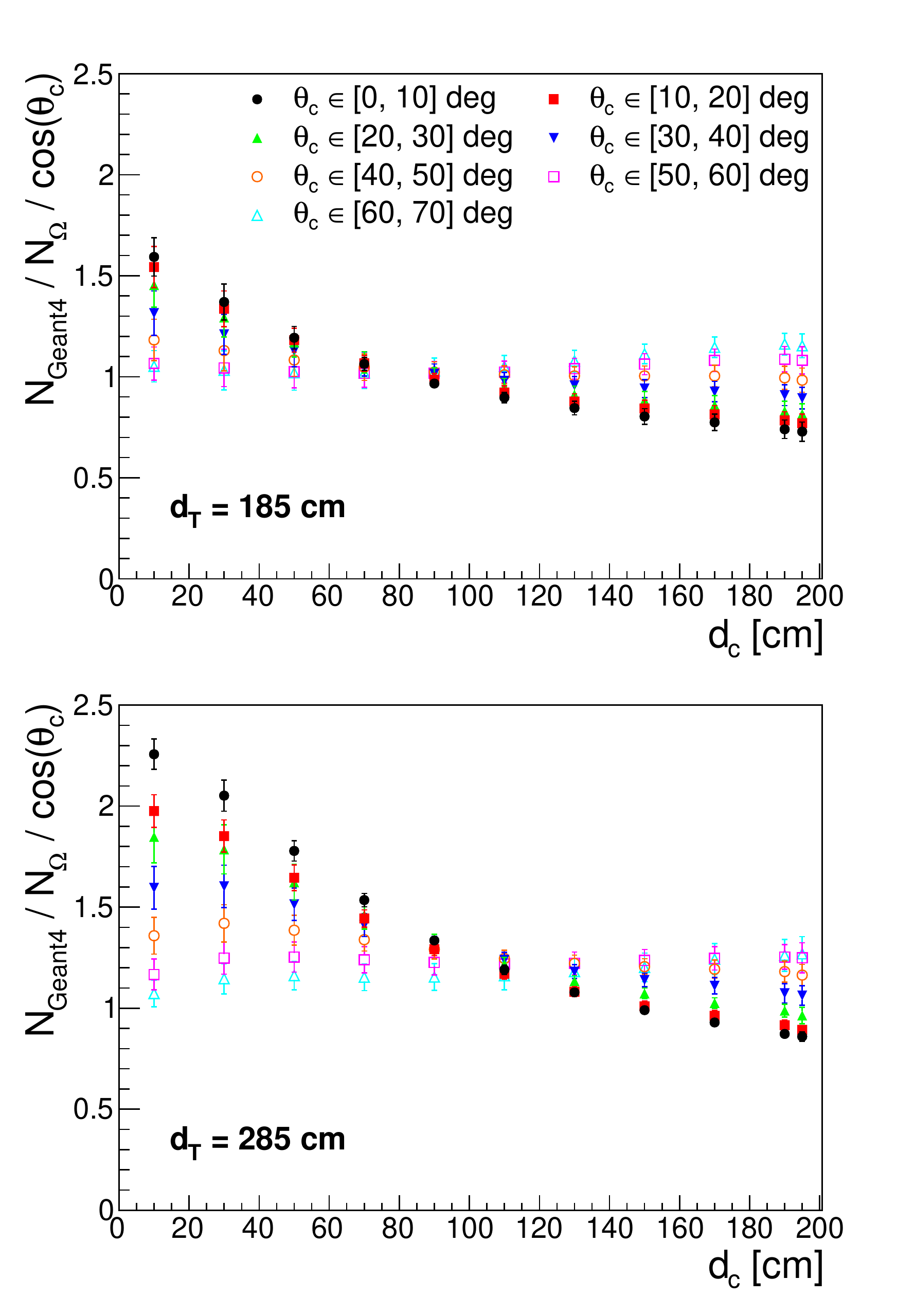}
	\caption{Examples of the border effect corrections required for the reflected light model in two different regions of the SBND-like detector geometry.}
	\label{fig:vis_SBND_border_corrections}
\end{figure}

Examples of sets of border effect corrections for the SBND-like detector geometry are shown in Fig. \ref{fig:vis_SBND_border_corrections} for two cylinders defined by different values of $d_T$. As before, the corrections are taken as the ratio between the amount of light seen in full simulation in Geant4 compared with the prediction from the geometric model. The equivalent corrections for the DUNE-like detector geometry are shown in \ref{sec:appendix_visible}. The required corrective factors become significantly larger as $d_T$ increases and the scintillation is closer to the edges and corners of the detector volume. Additionally, the angular dependence becomes more significant and larger offset angles of the PDs, as viewed from the bright-spot, become geometrically possible. The border effects are much more significant for the light reflected by wavelength-shifting foils compared with the direct light and larger corrective factors are therefore required.

Bringing the above effects together, the number of incident reflected light photons on each PD can be expressed as,
\begin{linenomath}
\begin{equation}
N_{\gamma, reflected} =N_{\Omega, reflected}  \times A(d_c, \theta_c, d_T) / cos(\theta_c),
\label{eq:Nvis_rec}
\end{equation}
\end{linenomath}
where $N_{\Omega, reflected}$ is the geometric prediction given by Eq.~\ref{eq:Nhits_vis} and $A(d_c, \theta_c, d_T)$ is the parameterized corrective factor accounting for PD position and border effects. This corrective factor depends only on the distance between the emission point and the bright-spot, the angle between the bright-spot and the PD, and the distance of the emission point from the center of the detector. As with the direct light, note that $N_{\gamma, reflected}=\Delta E \times S_{\gamma} \times Q_{abs} \times P(d_c,\theta_c)\times T(d_c,\theta_c)$ using the notation from Eq.~\ref{eq:first} for the reflected light.

\section{Predicting the Photon Arrival Time Distributions } \label{sec:parametrization}

The model developed in the previous section addressed only the prediction of the number of photons arriving at each PD but not the distribution of their arrival times. As described in Sections \ref{sec:production} and \ref{sec:RS}, the time of the scintillation light is dominated by the double-exponential distribution caused by the de-excitation of the two argon molecular dimer states.
In this section we describe a model to estimate the transport time $t_t(d,\theta)$, see Eq.~\ref{eq:tiempos}, that can affect the time distribution actually registered by the PDs. Analogous to Section~\ref{sec:rec_hits} we first develop a model for the direct light transport time and then use it as a starting point for a model describing light reflected off the cathode.

\subsection{Direct Light Time Parameterization}
\label{sec:parametrization_vuv}

\begin{figure}
\centering
\includegraphics[width=0.45\textwidth]{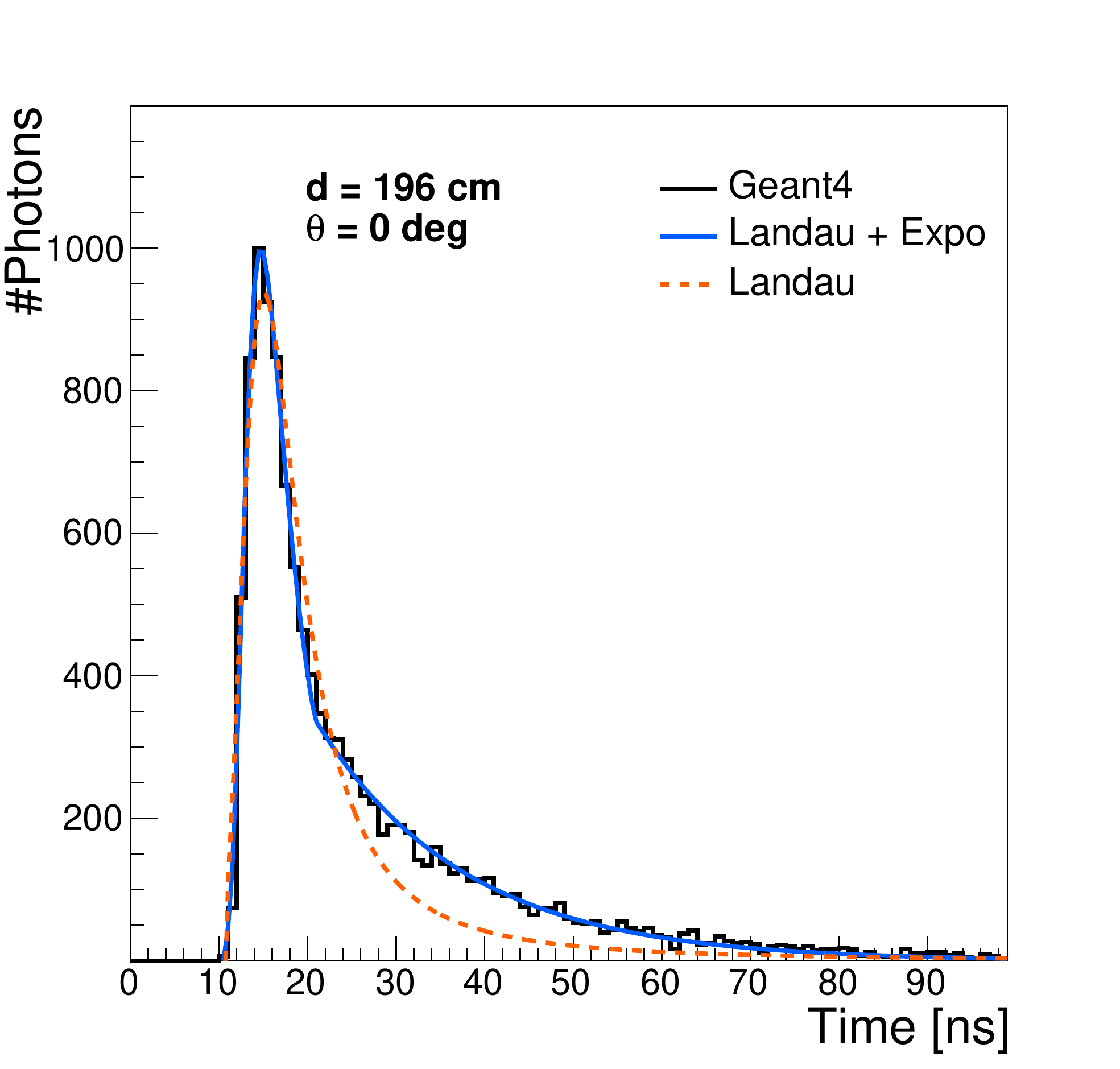}
\caption{Example of the distribution of direct light photon arrival times due to only transport effects together with the transport time models.} \label{fig:time_examples}
\end{figure}

\begin{figure}[!tbp]
\centering
\includegraphics[width=0.5\textwidth]{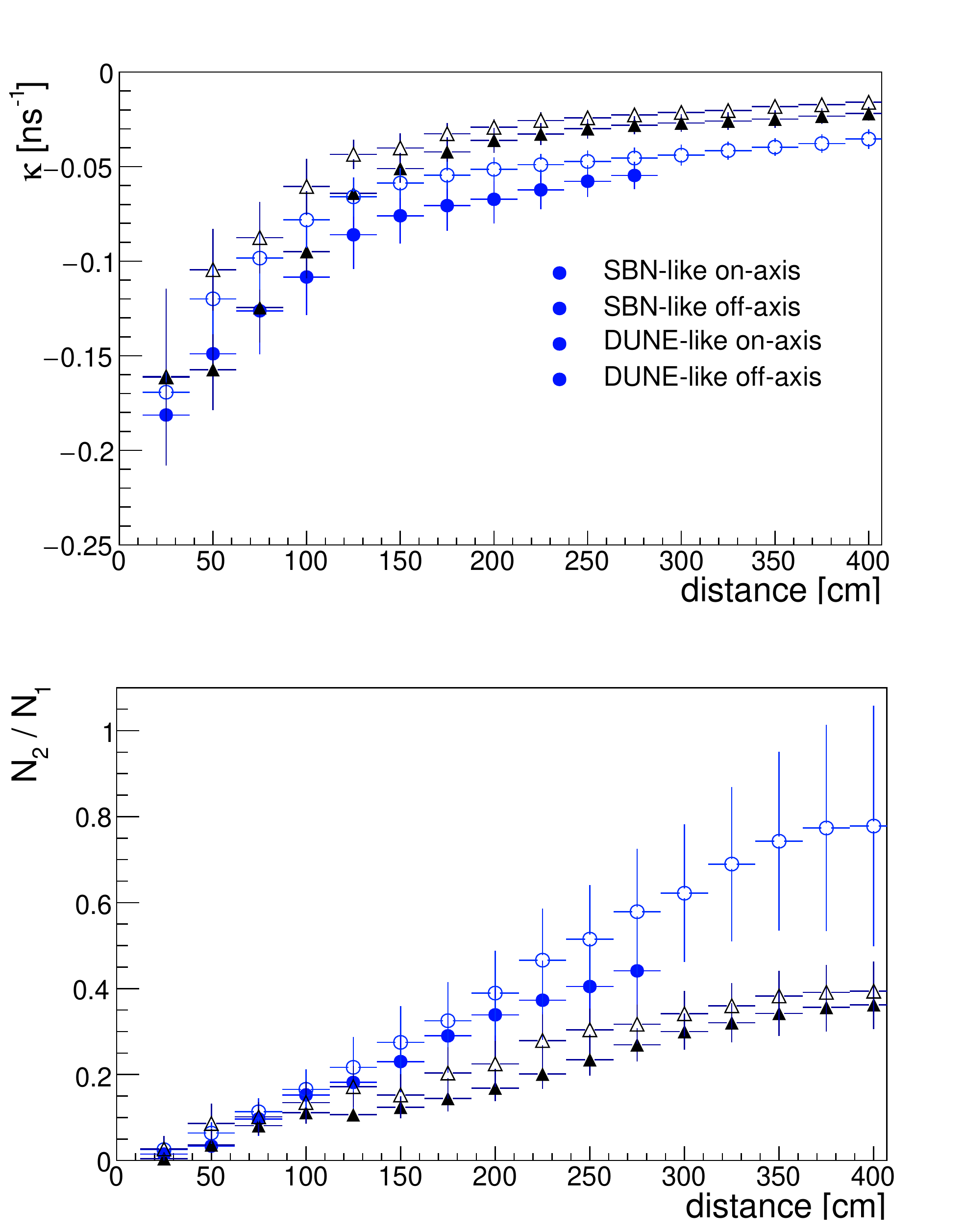}
\caption{Behavior of the Exponential component parameters of the direct light transport time model as a function of distance between the energy deposit and PD for the DUNE-like and SBND-like geometries.} \label{fig:time_model_Expo}
\end{figure}

The earliest arrival time of a photon on a particular PD can be predicted geometrically using the minimum distance that a photon must travel and the velocity of VUV light in LAr, shown in Fig. \ref{fig:op_prop}. A geometric calculation can provide the arrival time of the fastest possible photon, but does not account for other transport effects. A typical distribution of photon arrival times due to only transport effects can be seen in Fig.~\ref{fig:time_examples}. The distribution shows a prompt component followed by a long diffuse tail.
\begin{figure}[!tbp]
\centering
\includegraphics[width=0.5\textwidth]{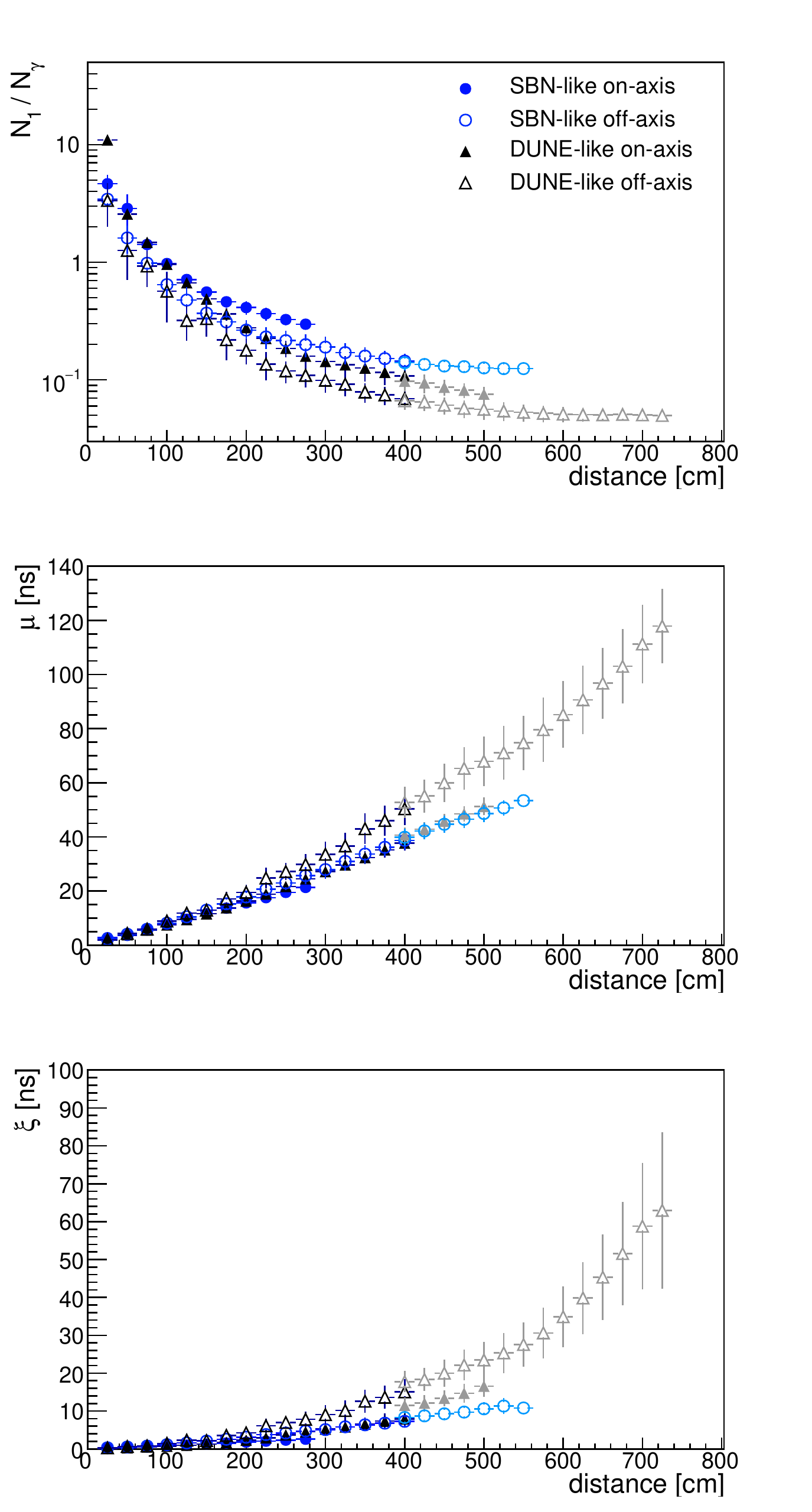}
\caption{Behavior of the Landau component parameters of the direct light transport time model as a function of distance between the energy deposit and PD for the DUNE-like and SBND-like geometries. The lighter grey and blue points denote the switch to using a simple Landau instead of the Landau + Exponential (for distances larger than $400$~cm).} \label{fig:time_model_Landau}
\end{figure}
We find empirically that for essentially all combinations of emission point and PD the distributions are of a similar nature and can be approximated by a combination of {\it Landau} and {\it Exponential} functions:
\begin{linenomath}
\begin{equation}
t_t(x)=\underbrace{N_1\,\frac{1}{\xi}\frac{1}{2\pi i} \int_{c - i\infty}^{c + i\infty} e^{\lambda s + s\,log\,s}\,ds}_{Landau} \,\,\,\,+ \underbrace{N_2\, e^{\kappa\,x}}_{Exponential},\\
\label{eq:t-model}
\end{equation}
\end{linenomath}
where $\lambda = \frac{x-\mu}{\xi}$, with $\mu$ and $\xi$ commonly referred as the landau {\it most probable value} and {\it width} parameters respectively, $\kappa$ is the slope of the exponential and $N_1$ and $N_2$ are normalisation constants. The resulting five parameters of the {\it Landau + Exponential} composite that describe a given time distribution are monotonic functions of the distance between the emission point and PD, provided we account for the incident angle. In the work described here we use two angular bins\footnote{An increase in the number of bins would result in greater accuracy, at the cost of increased computational time.}: on-axis with $\theta\in[0^{\circ}, 45^{\circ}]$, and off-axis with $\theta\in[45^{\circ}, 90^{\circ}]$.  Figures~\ref{fig:time_model_Landau} and \ref{fig:time_model_Expo} show the behavior of the model parameters for the two angle ranges in the two geometry cases: SBND-like (blue points) and DUNE-like (black points). The spread of the parameter values depends on the detector size: in a larger detector the signals are more scattered. For simplicity and because VUV photons are predominantly absorbed by all detector materials, we have neglected border effects in the model. 

\begin{figure}[!tbp]
\centering
\includegraphics[width=0.5\textwidth]{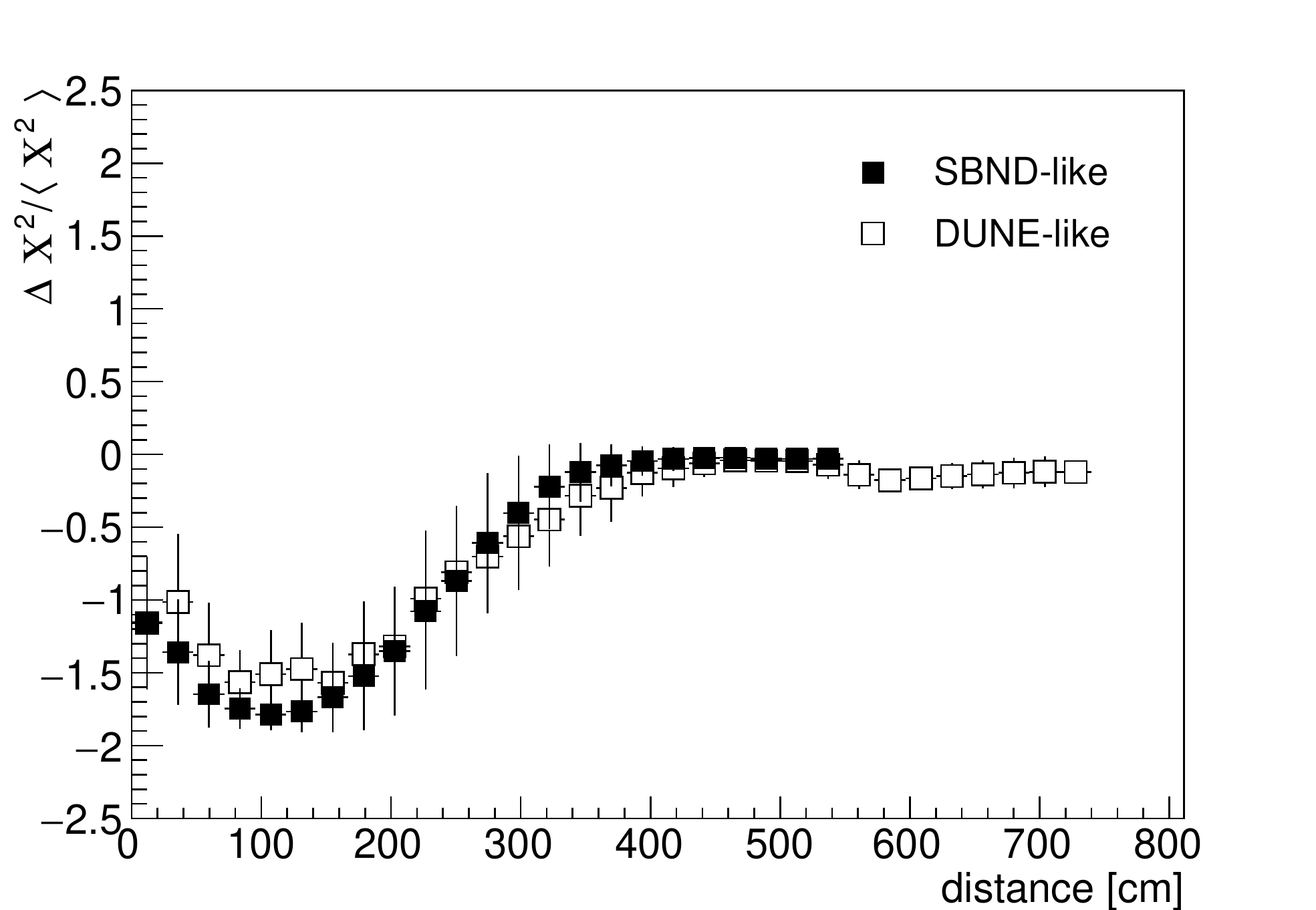}
\caption{Comparison of the relative difference of the $\chi^{2}$ for the two direct light transport time models: ``Landau + Exponential" vs ``Landau".} \label{fig:time_model_chis}
\end{figure}

At larger distances the long diffuse tail of the arrival time distributions tends to disappear and the shape can be described using only a Landau distribution. We perform a quantitative comparison of the accuracy of the two approaches, as a function of the distance, using the relative difference of the $\chi^{2}$ of both models. The result is shown in Fig.~\ref{fig:time_model_chis}. In both of the geometry cases we find similar results: the Landau + Exponential model describes the shape of our signals more accurately, but at larger distances the two models perform similarly. The distance at which the two models become equivalent depends very slightly on the detector size, but for both geometries under study has a value around $d = 400$\,cm. 
At longer distances the simpler Landau model can be used successfully instead of the Landau + Exponential one. 

During simulation we construct and then sample the probability distribution function (PDF) of the VUV photon arrival times for each PD using the parameters of the Landau + Exponential model. For computational reasons, we apply a cut-off at the 99$^{\text{th}}$-quantile when we sample the PDF of the transport time signal.

\subsection{Reflected Light Time Parameterization}

The transport time of photons arriving at the PDs as a result of a wavelength-shifting highly reflective layer on the cathode can be modelled using a similar approach. First, a geometric prediction of the transport time of earliest arriving photons is calculated. The fastest photons are most likely to travel along the path that minimizes the distance travelled at VUV wavelength, where the group velocity is slower. At visible wavelengths the photons propagate significantly faster due to the lower refractive index, as shown in Fig. \ref{fig:op_prop}. Figure~\ref{fig:vis_model_diagram} shows a diagram illustrating the most likely fastest path. The emitted VUV photons travel along the shortest path from the scintillation point to the cathode. There they are wavelength-shifted and re-emitted around the bright spot and take the shortest path to the PD. This simple model is able to predict the arrival time of the earliest photons.

The subsequent photons can be reflected from different regions of the wavelength-shifting foils and take very different paths to arrive at the PD. This results in a significantly broader distribution of their arrival times. We construct the model describing the visible photon arrival times at the PDs in three steps. We start by using the direct light Landau+Exponential model, described in Section~\ref{sec:parametrization_vuv}, to estimate the arrival time distribution of the VUV photons at the bright-spot on the cathode. We then add the time needed for a visible photon to propagate between the bright-spot and the PD in a straight line. Finally, we apply a parameterized smearing to the result to account for the multitude of longer paths that can be taken. We find empirically that the following smearing function effectively approximates the distribution,
\begin{linenomath}
\begin{equation}
t_s = t + (t-t_f)[\exp(-\tau \ln(x)) - 1],
\end{equation}
\end{linenomath}
where $t_s$ is the resulting smeared arrival time, $t$ is the un-smeared arrival time, $t_f$ is the fastest possible arrival time calculated geometrically, $\tau$ is a smearing factor and $x$ is a uniformly distributed random number between 0.5 and 1. This function keeps the earliest arrival times unchanged, but increasingly smears the photons arriving later. Additionally, a maximum time cut-off is applied to avoid an excessively long tail from the exponential distribution. 

We parameterize the smearing factor, $\tau$, and the cut-off time, $t_{max}$, in terms of the distance between the scintillation and the bright-spot, $d_c$, and the offset angle, $\theta_c$, between the bright-spot and the PD, as shown in Fig.~\ref{fig:vis_model_diagram}. The cut-off time is calculated as the time needed for $99.5$\% of Geant4 simulated photons to arrive. The $\tau$ parameter is determined by minimizing the difference between the smeared arrival time distribution calculated by the model and the distribution generated using Geant4. 

\begin{figure}
	\centering
	\includegraphics[width=.5\textwidth]{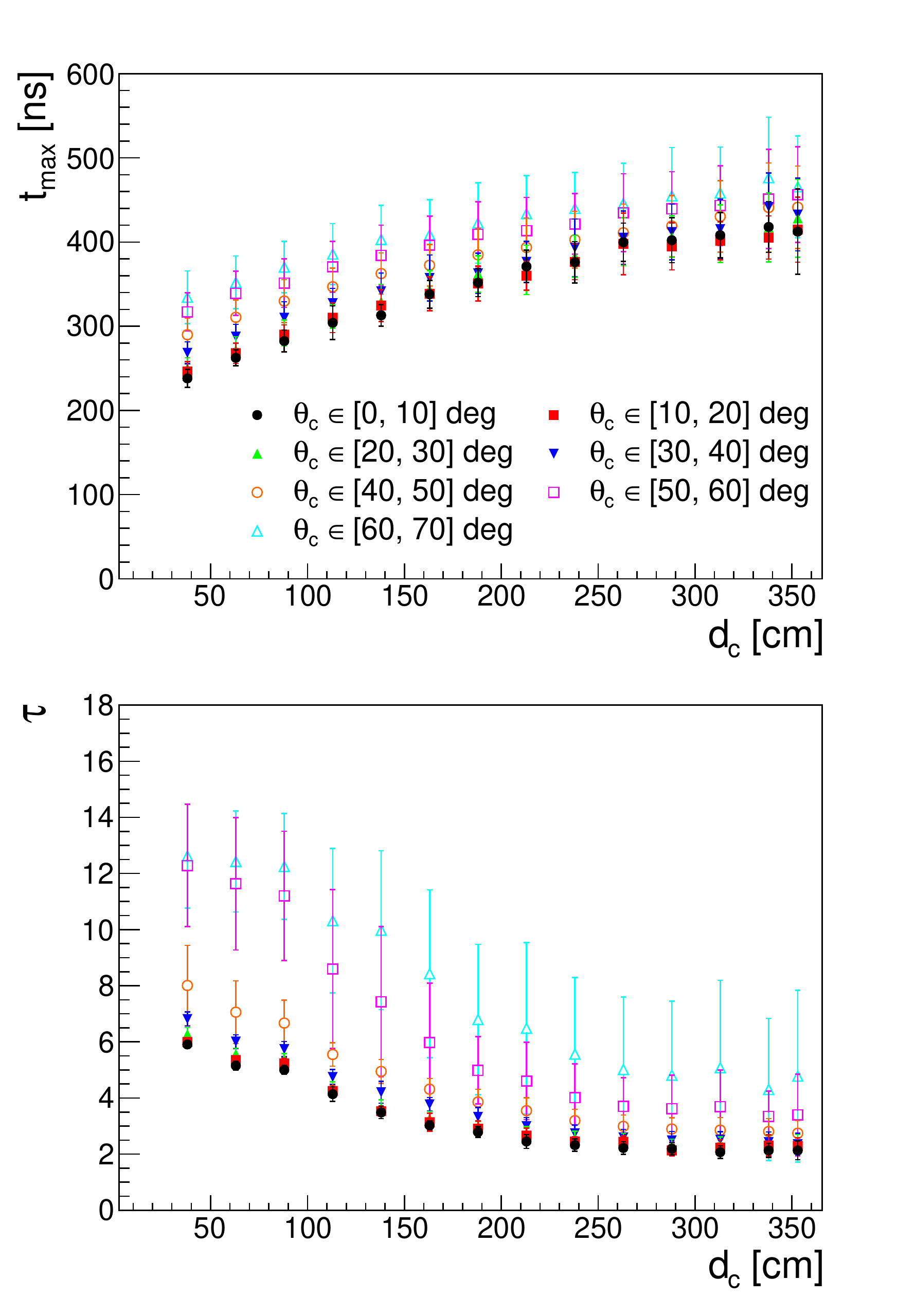}
	\caption{Reflected light transport time model cut-off time (top) and smearing parameter (bottom) in the central region of the DUNE-like detector geometry.} 
	\label{fig:vis_dune_time_pars}
\end{figure}

Unlike with the direct light transport time model, it is important to account for border effects such as reflections off the detector walls since they are highly reflective for visible photons. We use a similar approach to Section \ref{sec:NHits_vis}, creating sets of smearing parameters at different distances from the center of the detector, $d_T$. These sets can then be used to calculate the smearing parameters for any location in the detector using interpolation. An example set of the parameterized cut-off times and $\tau$ smearing factors is shown for the DUNE-like geometry in Fig. \ref{fig:vis_dune_time_pars}. An equivalent example for the SBND-like geometry is shown in \ref{sec:appendix_visible}. 

We observe that the cut-off times become larger with the size of the detector. This can intuitively be explained by the longer distances the photons need to travel before reaching the PDs, including many paths where they are reflected off the detector walls. 
The angular dependence of the cut-off time is relatively small, with a significant overlap between bins. The $\tau$ parameter is more dependent on the angle. This effect again grows with detector size and is much more prominent for the DUNE-like case. 

\section{Validation and Performance} \label{sec:validation}

To validate our model we test its performance against the results
of a full simulation of the scintillation light in Geant4. For this test we use a sample of points generated in an analogous manner to that from Section~\ref{sec:simulation_framework} but shifted by several centimetres in random directions to test how the model works in the interpolated areas. We also compare the performance of our model with that of optical lookup libraries and give an example of applying the model to a realistic event. 

\subsection{Predicting the number of detected photons}

\subsubsection{Direct Light}\label{sec:validation-vuv}

\begin{figure}
\centering
\includegraphics[width=.48\textwidth]{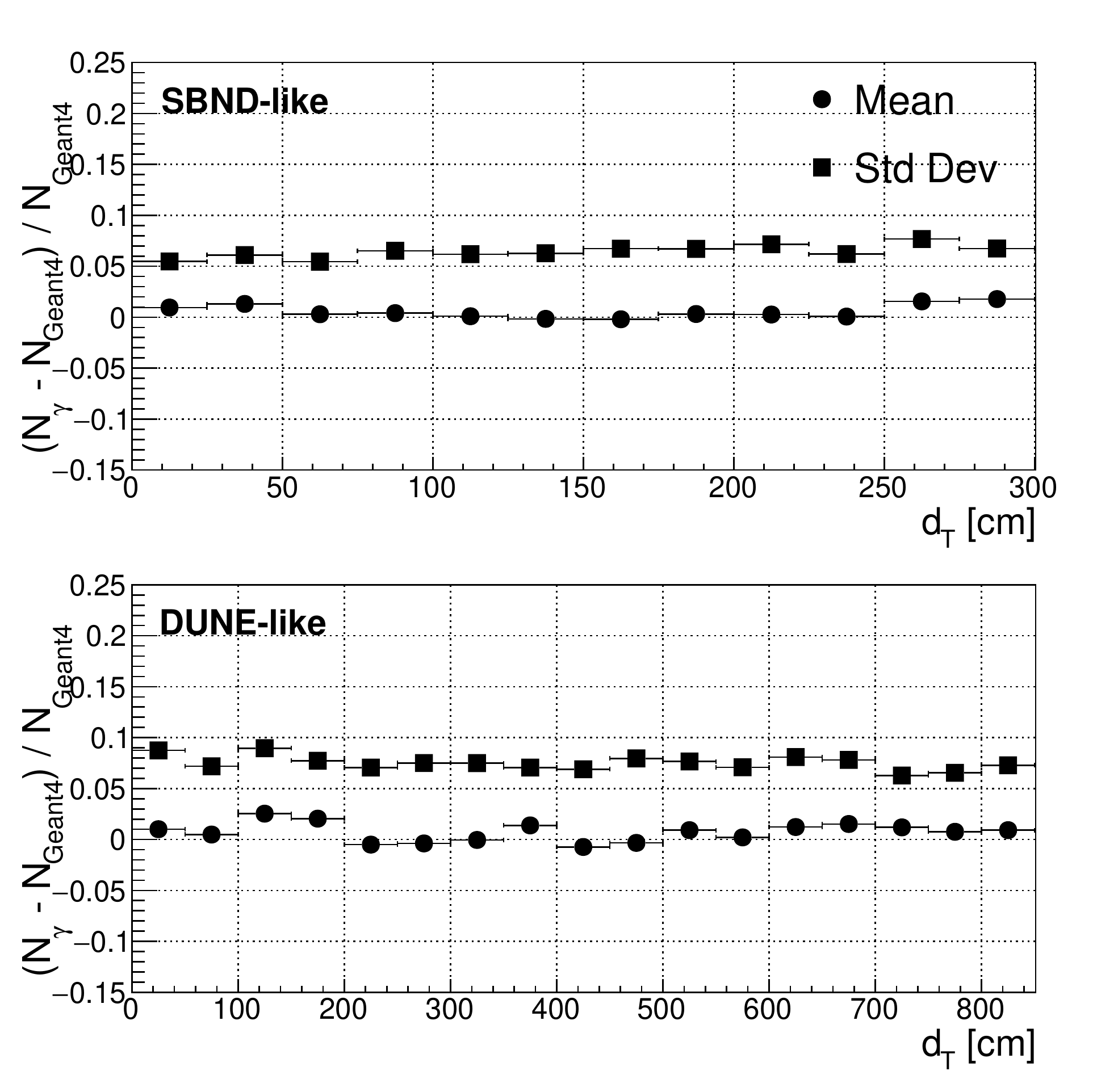}
\caption{Resolution of the direct light semi-analytic model as a function of the distance to the PD-plane center, $d_T$.
} 
\label{fig:res_vuv_radial_both}
\end{figure}

\begin{figure}
\centering
\includegraphics[width=.48\textwidth]{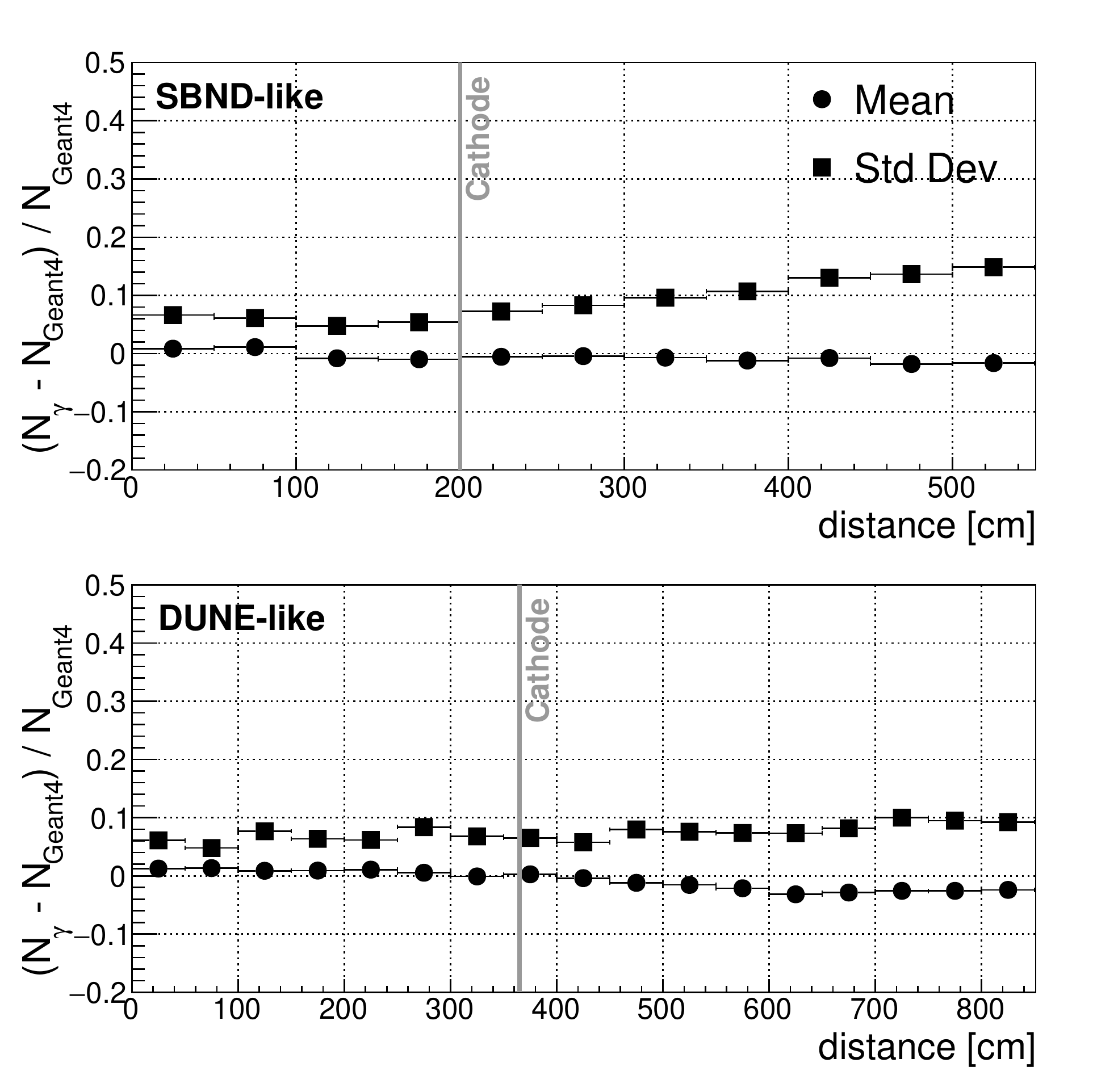}
\caption{Resolution of the direct light semi-analytic model as a function of the distance between the scintillation emission and the PD. The position of the cathode is illustrated for both geometries by the grey lines.} 
\label{fig:res_vuv1_both}
\end{figure}

The resolution obtained with the direct light semi-analytic model as a function of $d_T$ is shown in Fig.~\ref{fig:res_vuv_radial_both} for the SBND-like and DUNE-like geometries. We obtain an unbiased estimation of the number of VUV photons arriving to our PDs in both geometries and for all values of $d_T$. We also find the global resolution to be better than 10\%, independent of $d_T$. Figure \ref{fig:res_vuv1_both} shows the performance as a function of the distance between the scintillation emission and the PD. The resolution worsens slightly with distance, ranging from 5-15\% as we move from the closest to the farthest PDs. In each case, the performance is worst at the distances significant larger than the maximum drift distance (grey line) beyond which all PDs are off-axis. These PDs, however, are a minor contribution to the overall light signal of a physics event and do not significantly affect the overall resolution. 

\begin{figure}
	\centering
	\includegraphics[width=.48\textwidth]{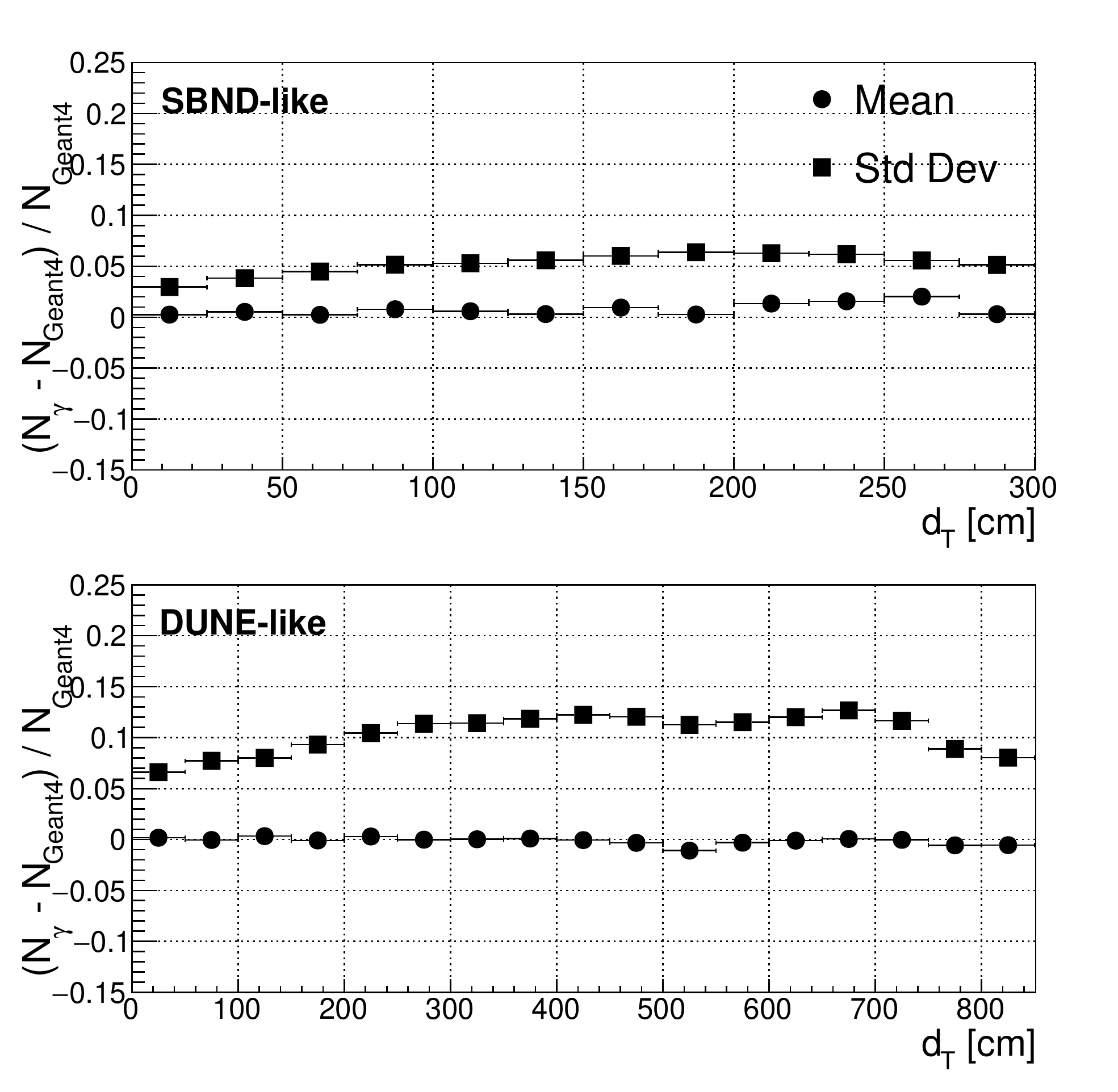}
	\caption{Resolution of the reflected light semi-analytic model as a function of the distance to the PD-plane center, $d_T$.
	}
	\label{fig:vis_resolution_profile}
\end{figure} 

\subsubsection{Reflected Light}
The resolution obtained with the reflected light semi-analytic model as a function of $d_T$ is shown in Fig. \ref{fig:vis_resolution_profile} for the SBND-like and DUNE-like geometries. 
The model performs well throughout the entire detector volume in both cases. It has a resolution better than $10\%$ in the SBND-like geometry and better than $15\%$ in the DUNE-like geometry, with minimal bias in each case. For both geometries, the resolution is best in the central region of the detector, at small $d_T$, where the effects of the borders are smallest. It then degrades slightly at larger $d_T$ as the border effects become more substantial and complex. The performance of the model in the DUNE-like case is poorer than for the SBND-like case due to the larger number of possible positions within the detector and larger number of different PDs for each $d_T$ and $\theta_c$ bin, especially at larger angles. This results in greater uncertainty and spread in the corrective factors required, as seen in Fig. \ref{fig:vis_centre_corrections}. Additional plots showing this effect can be found in \ref{sec:appendix_visible}.

\subsection{Predicting the photon arrival time distributions}

\subsubsection{Direct Light}

The performance of the direct light model at predicting the time of the earliest arriving photon, $t_0$, is shown in Table \ref{tab:timingres} for the SBND-like and DUNE-like geometries. In both cases $t_0$ is predicted with a resolution better than $0.5$ ns and with minimal bias. This resolution is smaller than the sampling of the PD electronics in current and upcoming LArTPC detectors, as described in Section \ref{sec:detection}.   

\begin{figure}[]
	\centering
	\includegraphics[width=.5\textwidth]{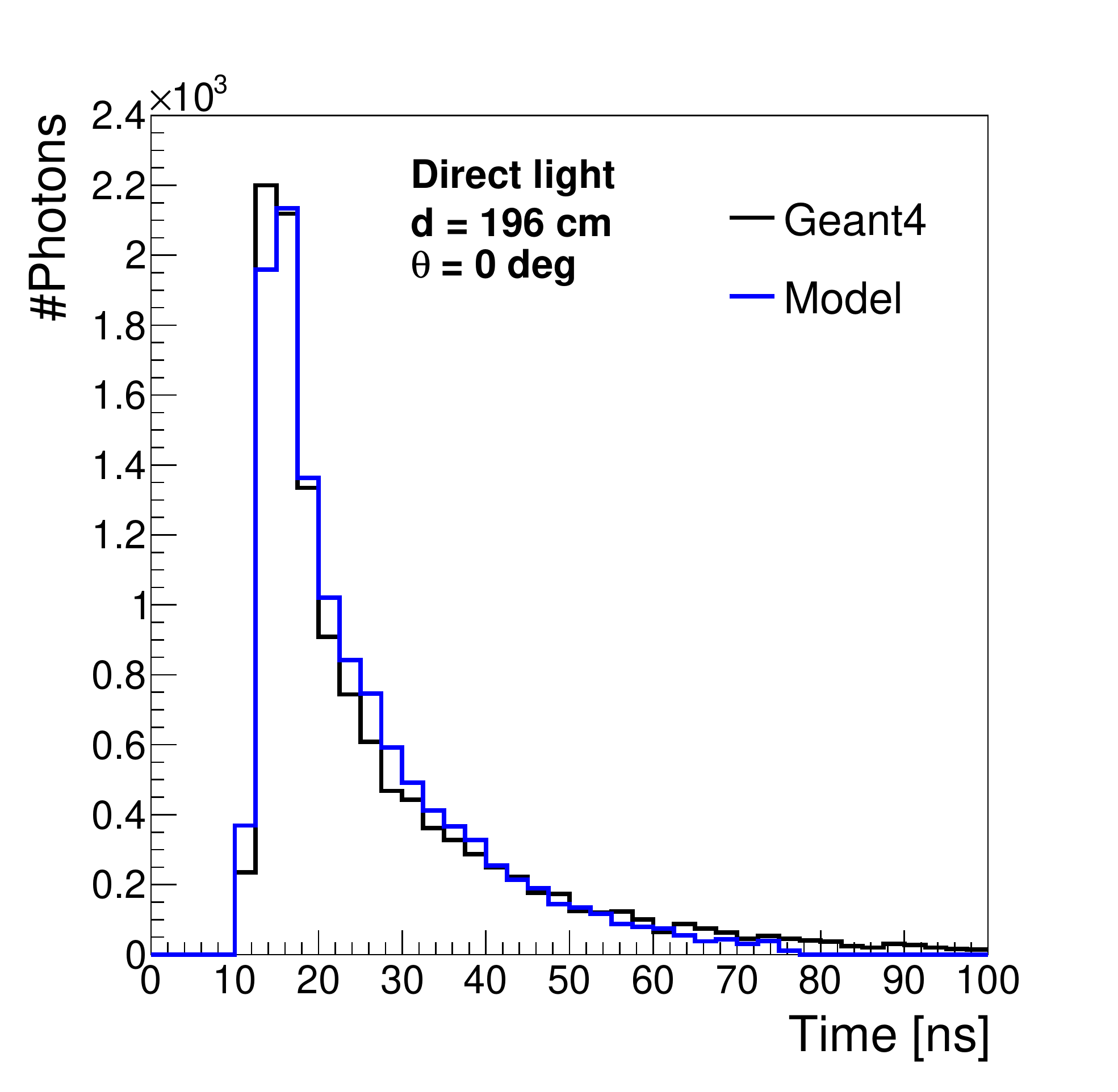}
	\caption{Example of the performance of the direct light transport time model compared with simulation in Geant4 in the SBND-like detector geometry.} 
	\label{fig:vuv_arrival_time_distribution}
\end{figure}

An example comparison between the direct light photon transport time distribution predicted by the model and simulation in Geant4 is shown in Fig. \ref{fig:vuv_arrival_time_distribution} for the SBND-like detector geometry. The distribution of the photon arrival times is accurately predicted, except for a very slight tail-offset between the distribution from Geant4 and the model. This is due to the example $\theta$ lying at the extreme edge of the parameterized $\theta\in[0^{\circ}, 45^{\circ}]$ angular bin. This offset could be reduced by increasing the number of angular bins used in the parameterization and by using variable bin sizes, with higher density in less linear regions.  

\begin{table}[h]
    \centering
    \begin{tabular}{|c|c|c|c|c|}
      \hline
        & \multicolumn{2}{c|}{SBND-like} & \multicolumn{2}{c|}{DUNE-like} \\ \hline
        model & mean & std dev & mean & std dev \\ \hline
     VUV: $\Delta t_0$ [ns] & $-0.2$  & $0.2$ & $0.0$ & $0.3$  \\ 
     Visible:  $\Delta t_0$ [ns] & $0.0$ & $0.3$ & $0.3$ & $0.9$  \\
        \hline
    \end{tabular}
    \caption{Resolution of the photon transport time model prediction of the earliest arriving photon time for the direct and reflected light in the SBND-like and DUNE-like geometries. In each case, $\Delta t_0~=~t_{0, Geant4}~-~t_{0, model}$. The uncertainties on the mean and standard deviation are negligible.}
    \label{tab:timingres}
\end{table}
\subsubsection{Reflected Light}

The performance of the reflected light model at predicting time of the earliest arriving reflected photon, $t_0$, is shown in Table \ref{tab:timingres} for the SBND-like and DUNE-like geometries. In the SBND-like geometry, $t_0$ is predicted with a resolution better than 0.5\,ns and without bias. In the DUNE-like geometry the performance is slightly worse, however the model still predicts $t_0$ with a resolution better than 1\,ns and minimal bias. As with the direct light model, these numbers are well inside the timing resolution of the PD electronics in typical LArTPC detectors.

\begin{figure}[]
	\centering
	\includegraphics[width=.5\textwidth]{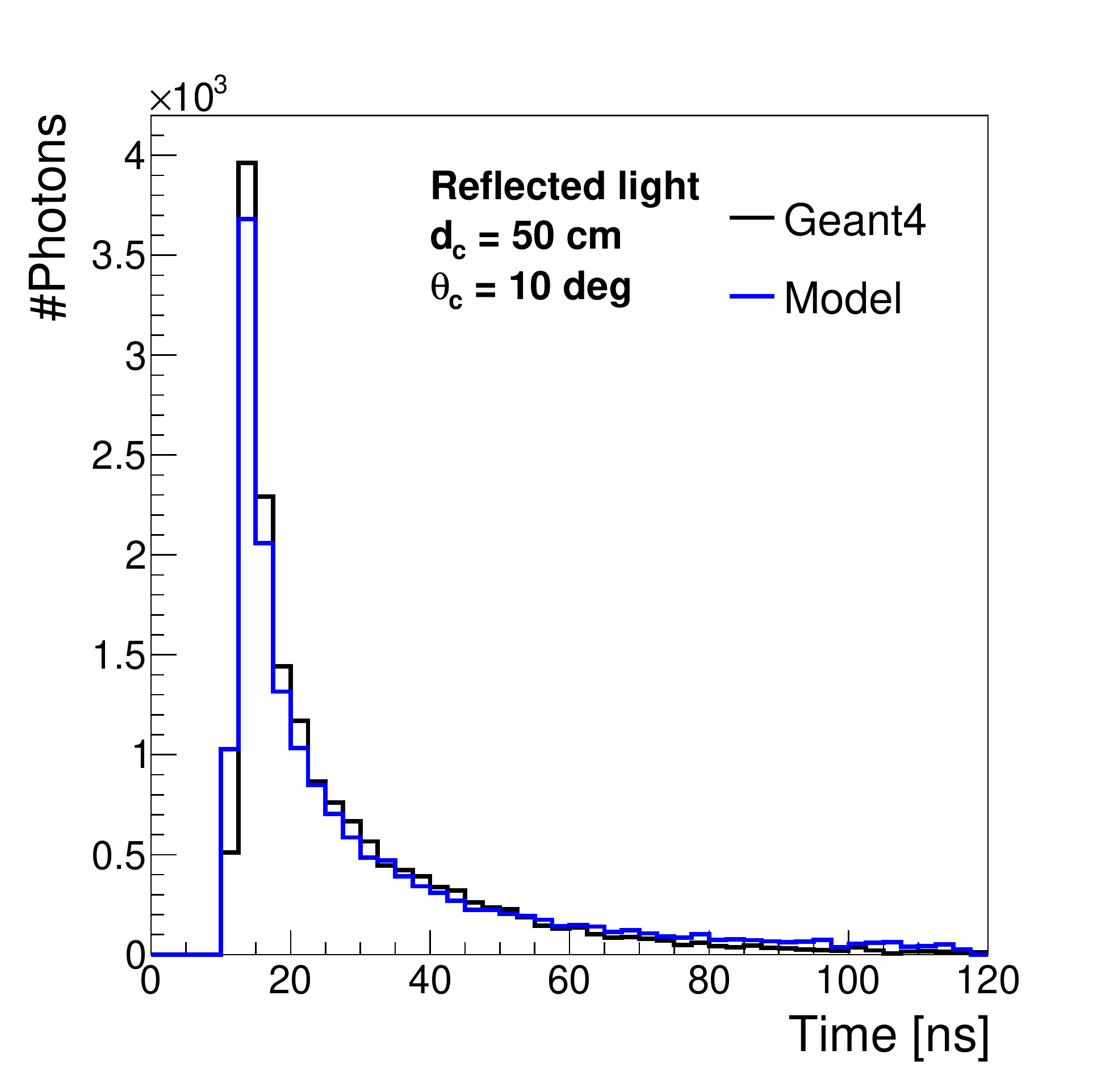}
	\caption{Example of the performance of the reflected light transport time model compared with simulation in Geant4 in the SBND-like geometry.} 
	\label{fig:visible_arrival_time_distribution}
\end{figure}

An example comparison between the reflected light photon transport time distributions predicted by the model and simulation in Geant4 is shown in Fig. \ref{fig:visible_arrival_time_distribution} for the SBND-like detector geometry. The model accurately predicts the arrival time of the earliest photons and provides a reasonable approximation of their overall distribution. The model slightly underestimates in the first part of the tail of the distribution and overestimates towards the end of it. We found that this behavior most prominently affects off-axis PDs, which see substantially less light than those closer to the energy deposit, resulting in a relatively small overall impact.

\subsection{Comparison with lookup libraries}

An important consideration is how the performance of the model developed here compares to that of the lookup library method commonly used in neutrino LArTPCs. We perform this test for both the SBND-like and DUNE-like detector geometries. To directly compare performance we generated dedicated lookup libraries with the same total number of photons used to train our model, see Table~\ref{tab:lookuplibraries} for details. We used a uniform voxel size throughout the detectors and a uniform distribution of photons/voxel\footnote{This is common practice in generating optical lookup libraries. However, we note that varying the voxel size or the number of photons/voxel could improve the performance compared to the results shown here. This would likely require a separate optimization process.}. For completeness we also generated a ``Hi-Res" version of the SBND-like lookup library to compare performance with a larger number of photons/voxel. The library generation takes approximately between 90\,s and 170\,s per $1\times10^6$ photons simulated, depending on the position in the detector.

\begin{figure}
	\centering
	\includegraphics[width=.5\textwidth]{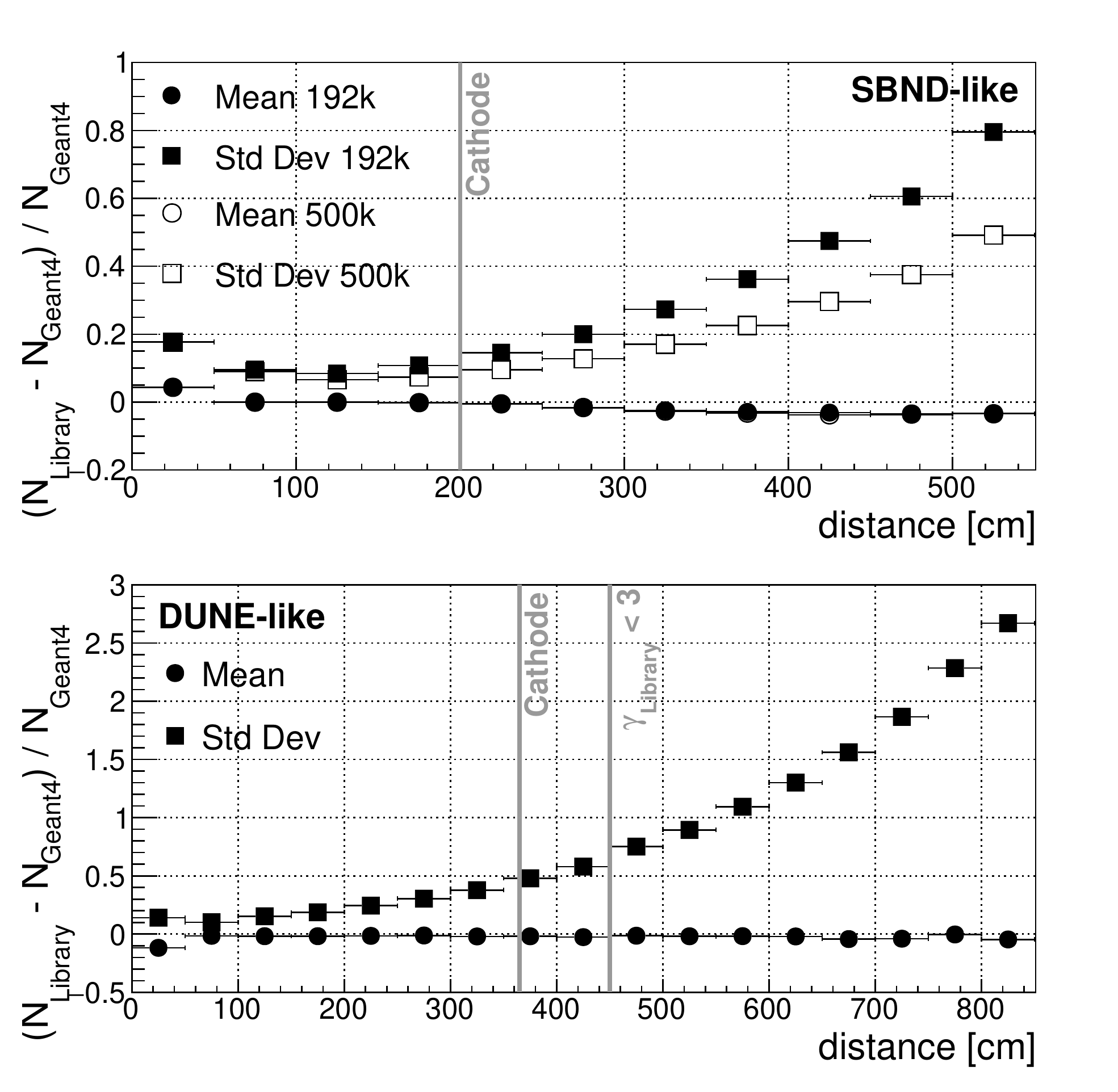}
	\caption{Performance of the lookup library method for the SBND-like and the DUNE-like geometries in the estimation of the number of direct light photons as a function of the distance between the scintillation emission and the PD. The black points in both plots were obtained using the same total number of photons to train the semi-analytic model shown in Fig.~\ref{fig:res_vuv1_both} (note the different axes). In the top plot the white points represent a lookup library generated with an increased total number of photons resulting in 500k$/$voxel. In the bottom plot a vertical line line indicates the distance beyond which the majority of the lookup library predictions are based on samples of less than 3 photons, which results in large fluctuations in the predictions of the library. In both plots, the position of the cathode is also illustrated by vertical lines.} 
	\label{fig:vuv_merged_resolution_library}
\end{figure}
\begin{table}[]
    \centering
    \begin{tabular}{|c|c|c|c|c|}
    \hline
    Library    & \makecell{Total Phot.} &  \makecell {Phot. per \\Voxel} &  \makecell{Voxel Size \\\text{[cm$^3$]}} & \makecell{Size\footnote{Note that the size of optical look-up libraries is proportional to the number of PDs, and that the SBND-like and DUNE-like geometries used in these studies have a factor of 4 fewer PDs than the real SBND and DUNE detectors.} \\\text{[MB]}} \\ \hline
    SBND-like  & \num{61.4e9}  & \num{192e3} & 5$\times$5$\times$5 & 390 \\
    DUNE-like    & \num{353.5e9} & \num{158e3} & 5$\times$5$\times$11 & 826 \\
    \makecell{SBND-like \\ Hi-Res} & \num{159.9e9} & \num{500e3} & 5$\times$5$\times$5 & 499 \\ \hline
    \end{tabular}
    \caption{Parameters of the lookup libraries generated to compare with the numerical model. Except for the ``Hi-Res" case, the total number of photons corresponds to the number of photons used to train the model presented in this work.}
    \label{tab:lookuplibraries}
\end{table}

The results of the comparison for the direct, VUV light, model can be seen in Fig.~\ref{fig:vuv_merged_resolution_library} for the two geometries under study. We compare these plots to Fig.~\ref{fig:res_vuv1_both}, where the performance for the semi-analytic model is shown (note the different axes). We find that our model behaves significantly better than the lookup libraries in terms of both bias and standard deviation, especially at larger distances. This is at least partially a result of under-sampling of the lookup libraries, as shown by the improved performance of the ``Hi-Res" library in the SBND-like case. In the DUNE-like case the fluctuations are exacerbated by the fact that for distances larger than 450\,cm the severe under-sampling in photons/voxel at the library generation stage causes the majority of predictions to be based on samples of less than 3 photons per voxel-PD pair.
Additionally, at very short distances the lookup libraries suffer from a higher uncertainty due to the voxel size introducing discrete jumps in the predictions very close to the PDs. 
This second problem cannot be resolved by increasing the number of photons used per voxel, instead it requires reducing the size of the voxels or using a different approach altogether in this region.
\begin{figure}
	\centering
	\includegraphics[width=.5\textwidth]{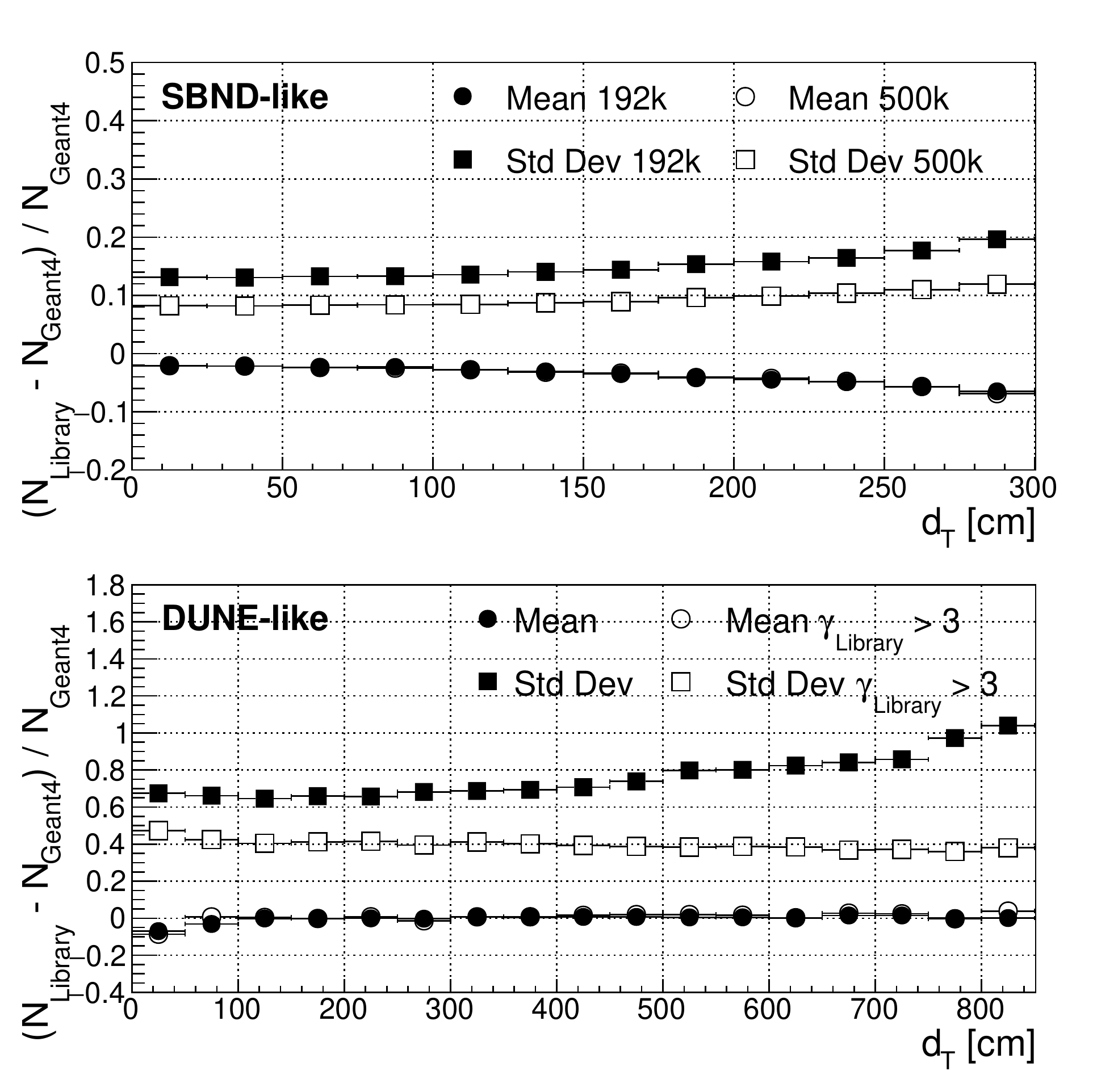}
	\caption{Performance of the lookup library method for the SBND-like and the DUNE-like geometries in the estimation of the number of reflected light photons as a function of the distance from the center of the PD-plane, $d_T$. The black points in both plots were obtained using the same total number of photons to train the semi-analytic model shown in Fig.~\ref{fig:vis_resolution_profile} (note the different axes). In the top plot the white points represent a lookup library generated with an increased total number of photons resulting in 500k$/$voxel. In the bottom plot the white points represent predictions generated only using voxel-PD pairs where the number of photons was larger than 3.} 
	\label{fig:visible_merged_resolution_both}
\end{figure}
\begin{figure*}
\centering
\includegraphics[width=\textwidth]{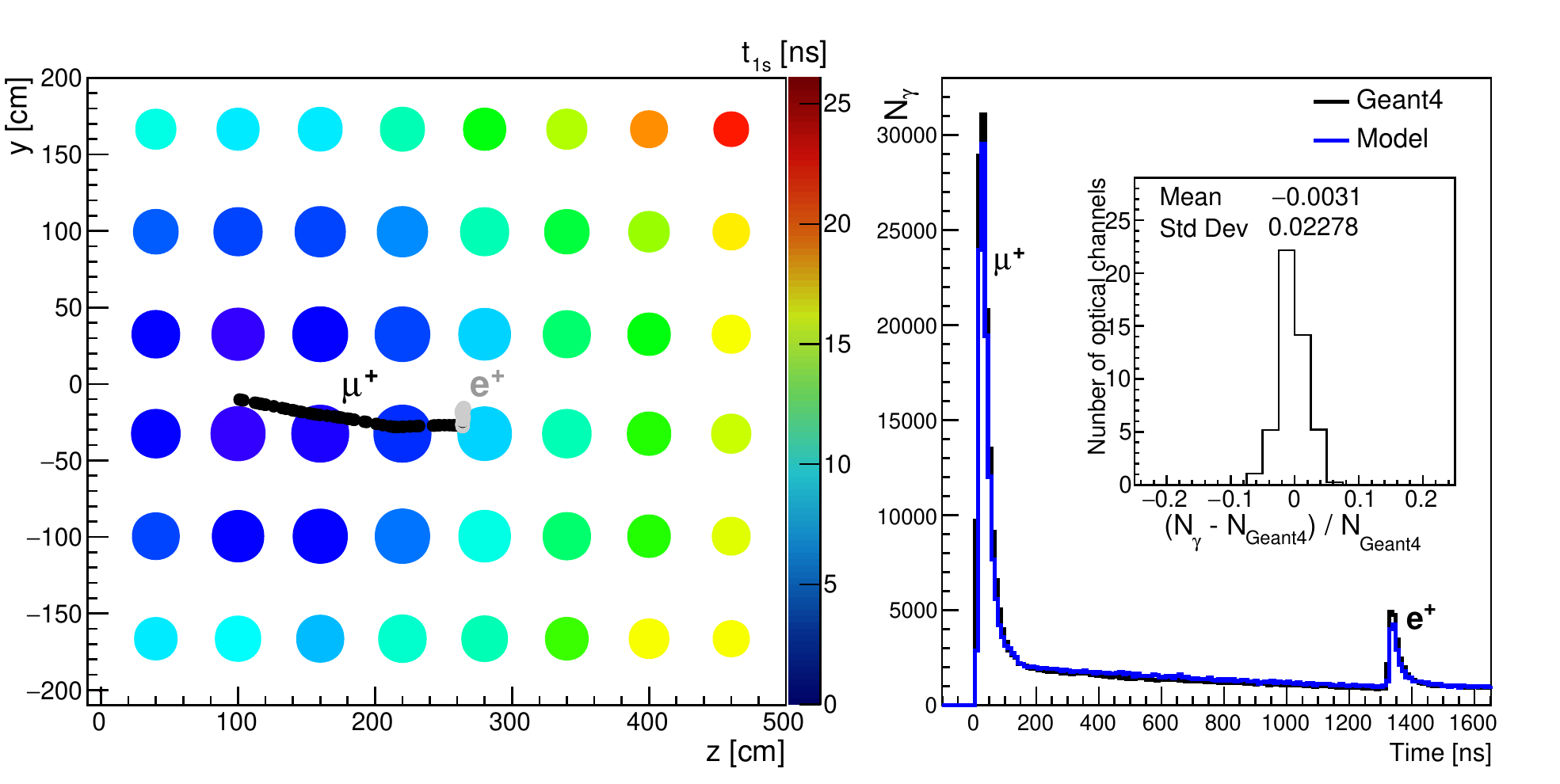}
\caption{
Event display of a stopping anti-muon ($\APmuon \rightarrow \APpositron \Pnue \APnum$) simulation. The left figure shows the charge (Geant4) and light (semi-analytic model) footprint projected on the PD-plane. Each circle represent a PD, where different colors indicate the starting time $t_{1st}$ of the signals (the anti-muon is entering from the left), and the size is proportional to the number of detected photons ($\propto log_{10}N_{\gamma}$). The right figure shows the summed Geant4 signal of all of the PDs overlapped with our model prediction, for comparison. We see excellent agreement between them, quantified by the resolution histogram of the light-model for this particular event.} 
\label{fig:display_example}
\end{figure*}

Figure~\ref{fig:visible_merged_resolution_both} shows the performance of the generated lookup libraries in predicting the number of reflected light photons. Due to the nature of modelling the reflected light we use $d_T$ as the variable instead of distance from the PD. We compare these plots to Fig.~\ref{fig:vis_resolution_profile}, where the performance for the semi-analytic model is shown. We observe that in the SBND-like geometry the lookup library method performs comparably to the semi-analytic model, especially if the ``Hi-Res" version is used, although a small under-prediction is observed in the regions of high $d_T$. In the DUNE-like case the under-sampling effects are so severe that the standard deviation of the lookup library prediction is much larger than for the semi-analytic method. The effect is again caused by many voxel-PD pairs where the prediction is made based on samples of a few photons. This could be mitigated by using a significantly higher number of photons/voxel to generate the lookup library.

Overall we find that the semi-analytic model performs significantly better than lookup libraries trained using the same number of photons. 

\subsection{Example Application to Realistic Events}

We have shown that our model works well for predicting the number of photons and their arrival times from point-like energy depositions. In simulations of particle detectors we more often deal with ``extended" objects such as tracks or showers. Our model can easily simulate these kinds of events using the paradigm used e.g. in Geant4, where particle trajectories are composed of discrete energy depositions called steps. To simulate realistic particle events we can apply our model to each of these steps and combine the results to obtain the simulation of the full particle trajectory. An example of this approach is shown in Fig.~\ref{fig:display_example}, where we present the results of simulating the scintillation light originating from an anti-muon track decaying into a Michel positron inside the SBND-like geometry. 

We also compare the prediction of the waveform observed by the PDs that our model makes to that of the full Geant4 simulation. We find excellent agreement for both the primary anti-muon scintillation peak and the secondary scintillation peak caused by the positron.

\section{Conclusions}

We have presented a new method to predict the number of scintillation light photons incident on photon detectors and their arrival times that can be used for simulations in large liquid argon neutrino detectors. Our model could also in principle be applied to any detectors constructed from materials where the Rayleigh scattering length is comparable to the volume size. Two scenarios were considered: VUV scintillation light that propagates directly to the photon detectors and scintillation light that is reflected off a wavelength-shifter coated highly reflective cathode. In each case, the models start with a prediction from pure geometric considerations, then corrections are applied for photon transport and border effects. For the prediction of the direct VUV light, we obtain a resolution better than 10\% in two different geometries: one SBND-like and one DUNE-like. For the reflected light, we obtain comparable performance in the smaller SBND-like detector and better than 15\% resolution in the larger DUNE-like detector.
In both scenarios, the prediction of the earliest photon arrival time provided by the models is within one nanosecond - better than the highest sampling used in liquid argon neutrino detectors to date. The method we propose is dramatically faster than the full Geant4 optical simulation and outperforms the currently used lookup library method when trained with the same number of fully simulated photons.
It can be used in any large scale liquid argon detector, as well as liquid xenon or xenon-doped argon detectors, with a simple tuning of the model parameters. 

\section{Acknowledgements}
The authors would like to thank Dr. Ettore Segreto and Dr. Flavio Cavanna for their extremely helpful comments and discussion which helped improve the manuscript.
This project has received funding from the European Union’s Horizon 2020 research and innovation programme under the Marie Sk$\l$odowska-Curie grant agreement N$^{o}$~754446 and UGR Research and Knowledge Transfer Found-Athenea3i; as well as from the Science and Technology Facilities Council (STFC), part of the United Kingdom Research and Innovation; and from the Royal Society UK awards: RGF\textbackslash EA\textbackslash 180209 and UF140089; as well as from Junta de Andalucía, Fondos FEDER (A-FQM-211-UGR-18), Junta de Andalucía (SOMM17/ 6104/ UGR).

\newpage
\appendix
\phantom{...}
\newpage
\section{Additional Tuning Plots: Direct Light Model} \label{sec:appendix_vuv}

In this section we show the remaining border effect tuning plots for the direct light model for both detector geometries.

\begin{figure} [hb]
  \includegraphics[width=.48\textwidth,keepaspectratio]{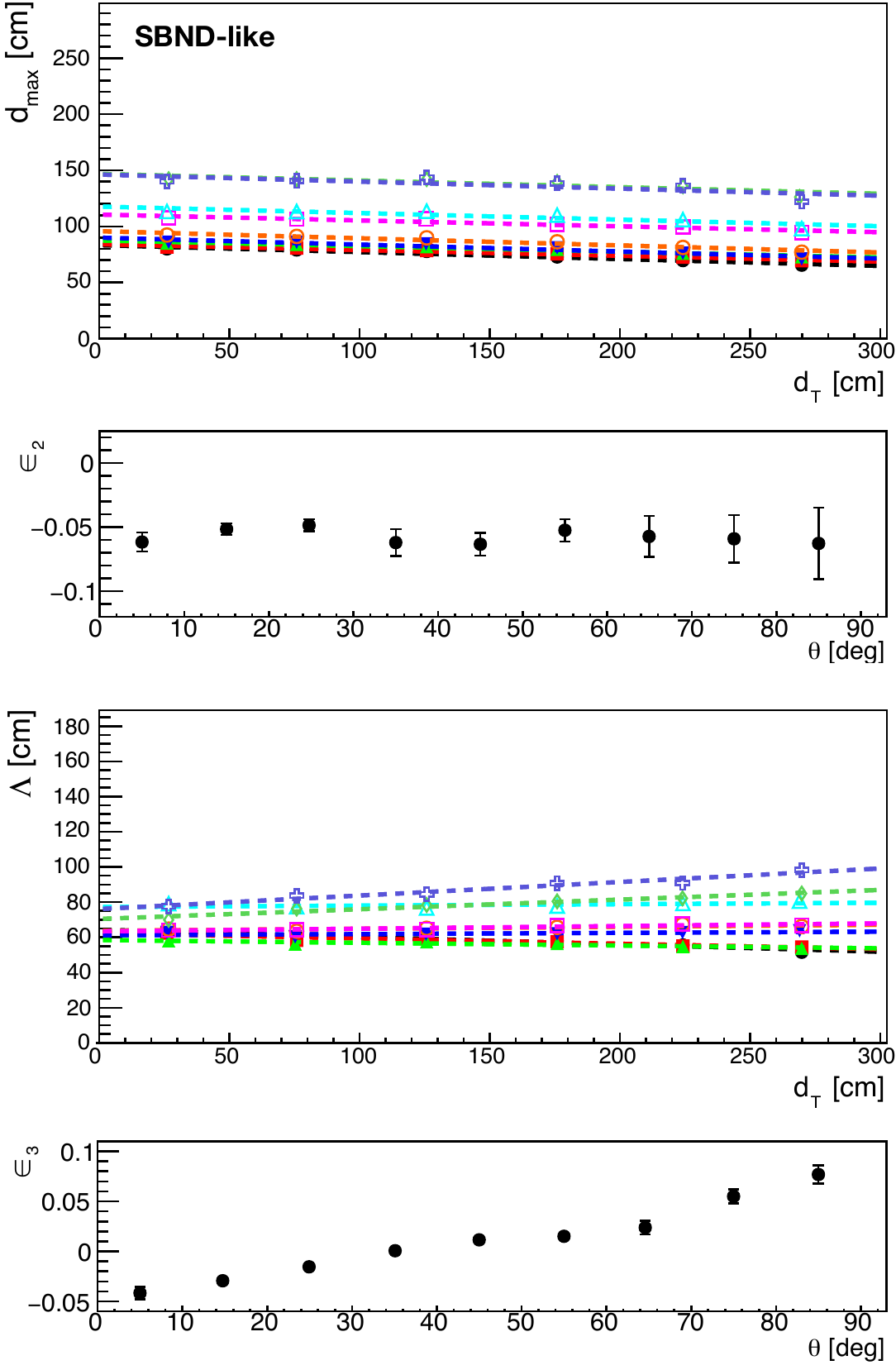}
  \caption{The upper panels show the $d_{max}$ and $\Lambda$  Gaisser-Hillas parameters dependency on distance to PD-plane center $d_{T}$ for the SBND-like geometry. The different colors represent different $\theta$ bins as shown in Fig.~\ref{fig:SBN_rainbow}. The lines represent the linear fit of the points. The lower panels show the slopes of the linear fits for the different offset angles.}
  \label{fig:border_slopes_SBN}
\end{figure}

\begin{figure} [ht]
  \includegraphics[width=.48\textwidth,keepaspectratio]{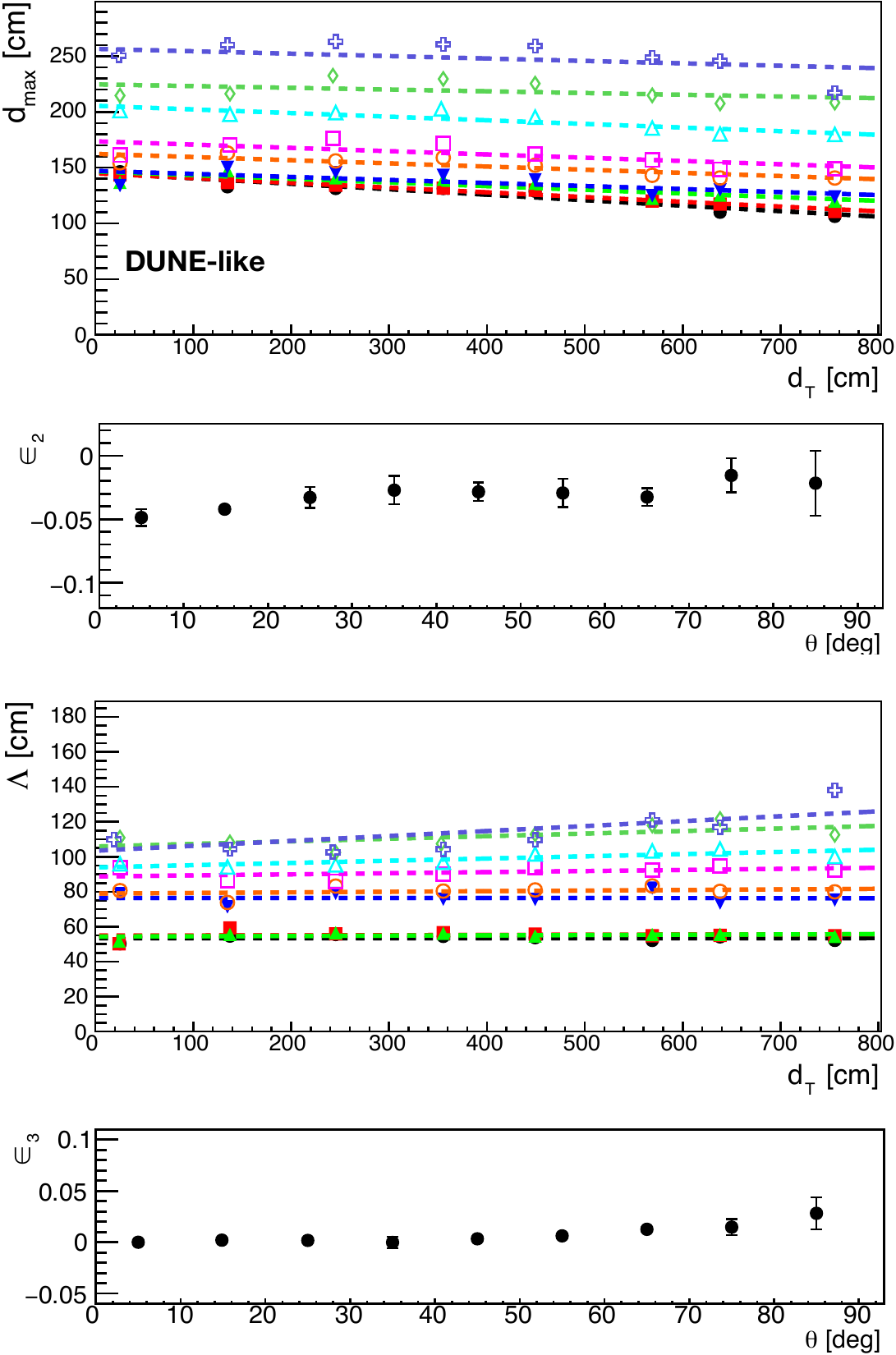}
  \caption{The upper panels show the $d_{max}$ and $\Lambda$  Gaisser-Hillas parameters dependency on distance to PD-plane center $d_{T}$ for the DUNE-like geometry. Different colors represent different $\theta$ bins as shown in Fig.~\ref{fig:SBN_rainbow}. The lines represent the linear fit of the points. The lower panels show the slopes of the linear fits for the different offset angles.}
  \label{fig:border_slopes_DUNE}
\end{figure}

Figures \ref{fig:border_slopes_SBN}~and~\ref{fig:border_slopes_DUNE} show the border corrections for the $d_{max}$ and $\Lambda$ parameters of the Gaisser-Hillas functions in Eq.~\ref{eq:GH_basic} for the SBND-like and DUNE-like geometries, respectively.

~

\newpage
~$\mbox{ }$
\newpage

\newpage
\section{Additional Tuning Plots: Reflected Light Model} \label{sec:appendix_visible}

\begin{figure}[ht]
	\centering
    \includegraphics[width=.5\textwidth]{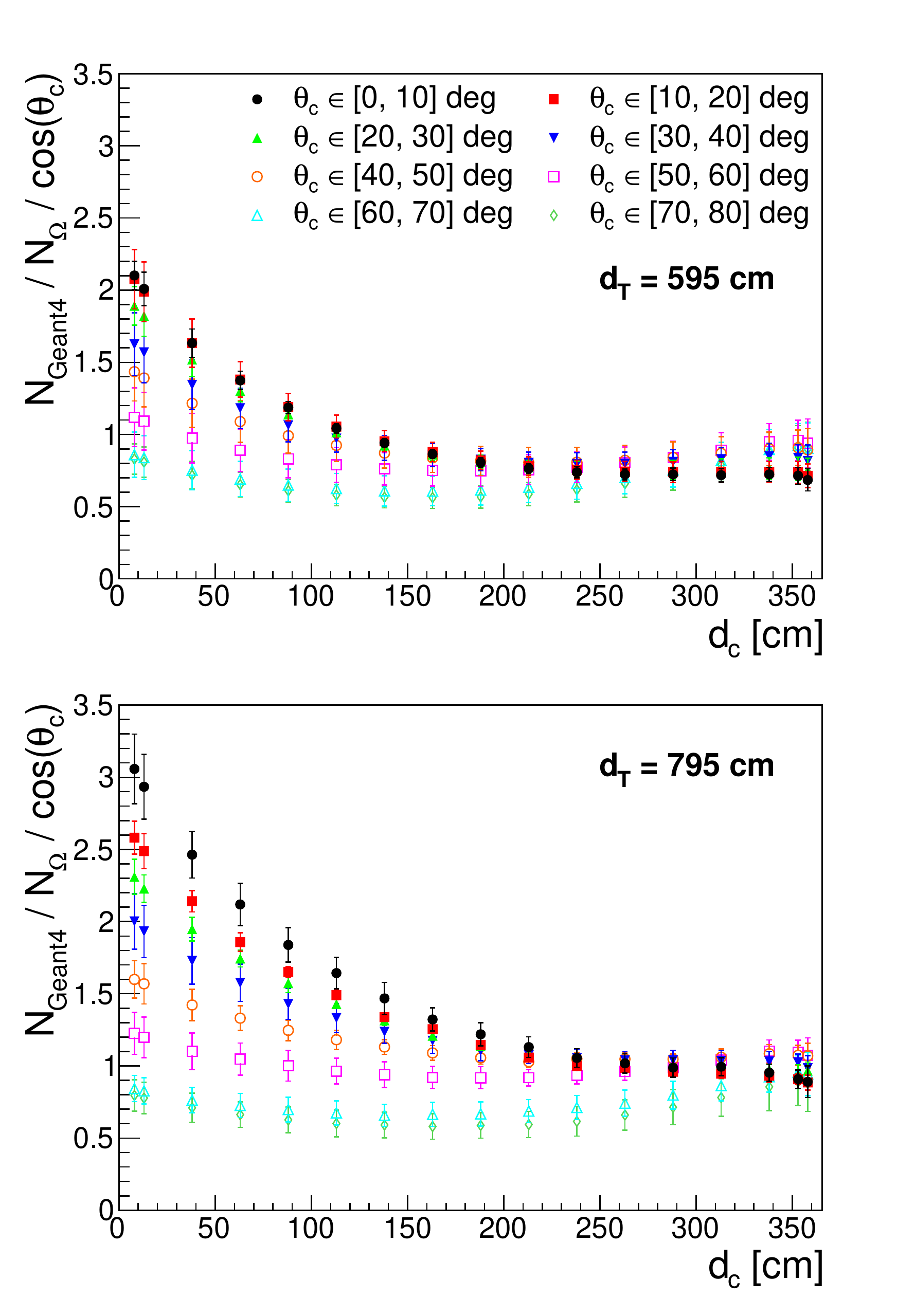}
	\caption{Examples of the border effect corrections required for the reflected light semi-analytic model in two different regions of the DUNE-like geometry.}
	\label{fig:vis_DUNE_border_corrections}
\end{figure}

In this section we show additional tuning plots for the reflected light model. Figure \ref{fig:vis_DUNE_border_corrections} shows examples of the reflected light semi-analytic model border corrections at two different $d_T$ for the DUNE-like geometry.

\begin{figure}
	\centering
	\includegraphics[width=.5\textwidth]{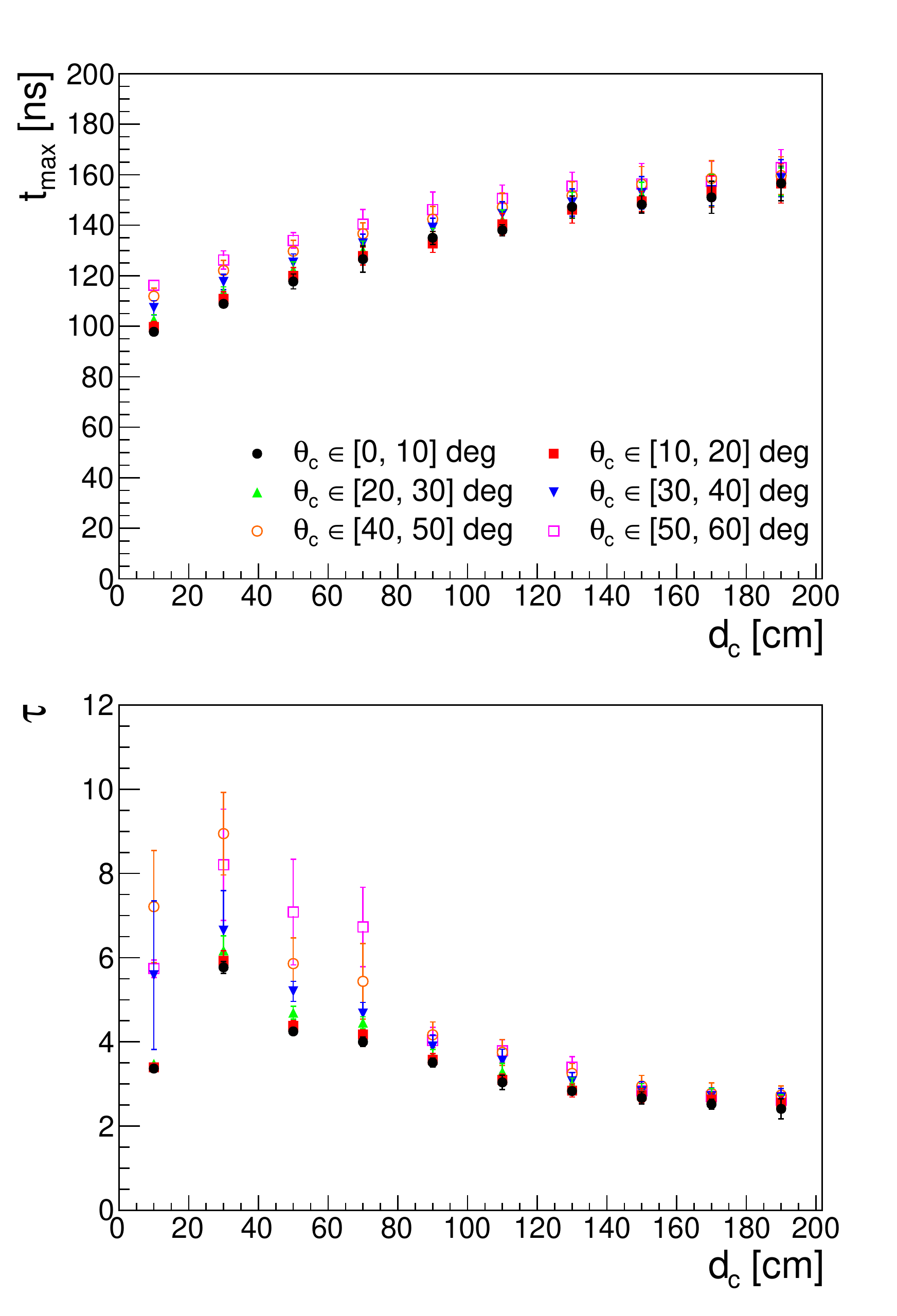}
	\caption{Reflected light transport time model cut-off time (top) and smearing parameter (bottom) in the central region of the SBND-like geometry.} 
	\label{fig:vis_sbnd_time_pars}
\end{figure}

Figure \ref{fig:vis_sbnd_time_pars} shows the reflected light transport time model cut-off times and $\tau$ parameters for the central region of the SBND-like geometry.

\clearpage

\bibliography{mybib}
\bibliographystyle{unsrt}

\end{document}